\documentclass{tlp}
\usepackage{aopmath}
\usepackage{url}
\usepackage[all]{xy}

\newcommand{\nat}{\mathrm{I\!N}}
\newcommand{\Expr}[2]{\textit{Expr$(#1,\mathcal{#2})$}}
\newcommand{\Term}[2]{\textit{Term$(#1,\mathcal{#2})$}}
\newcommand{\Bexpr}[2]{\textit{Expr$_{\bot}(#1,\mathcal{#2})$}}
\newcommand{\Bterm}[2]{\textit{Term$_{\bot}(#1,\mathcal{#2})$}}
\newcommand{\tterm}{\textit{tterm}}
\newcommand{\pterm}{\textit{pterm}}
\newcommand{\pexpr}{\textit{pexpr}}
\newcommand{\Sign}{\textup{\textbf{Sign}}}
\newcommand{\Sen}{\textit{sen}}
\newcommand{\Mod}{\textup{\textbf{Mod}}}
\newcommand{\Set}{\textup{\textbf{Set}}}
\newcommand{\Cat}{\textup{\textbf{Cat}}}
\newcommand{\Th}{\textup{\textbf{Th}}}
\newcommand{\True}{\textit{true}}

\newcommand{\lsem}{[\![}
\newcommand{\rsem}{]\!]}
\newcommand{\comp}{\mathop{\raisebox{.2ex}{$\scriptstyle\circ$}}}
\newcommand{\bigatop}[2]{{{\displaystyle #1}\atop{\displaystyle #2}}}
\newcommand{\bigfrac}[2]{\frac{\displaystyle #1}{\displaystyle #2}}

\newcommand{\ifcond}{\textup{\ \textbf{if}\ }}

\submitted{12 December 2003}
\revised{12 April 2005}
\accepted{5 January 2006}

\title{A comparison between two logical formalisms for rewriting}
\author[M. Palomino]{MIGUEL PALOMINO\thanks{Supported by a 
 postgraduate scholarship from the Spanish Ministry for Education,
 Culture, and Sports, and by the Spanish CICYT project AMEVA 
 TIC 2000--0701--C02--01. This work was completed during a stay of the
 author at the Department of Computer Science in the University of Illinois
 at Urbana-Champaign.} \\
 Departamento de Sistemas Inform\'aticos y Programaci\'on \\
 Facultad de Inform\'atica, Universidad Complutense de Madrid, Spain\\
 \email{miguelpt@sip.ucm.es}}

\begin{document}

\maketitle

\begin{abstract}
Meseguer's rewriting logic and the rewriting logic CRWL are two 
well-known approaches to rewriting as logical deduction that, despite
some clear similarities, were designed with different objectives.
Here we study the relationships between them, both at a syntactic
and at a semantic level.
Even though it is not possible to establish an entailment system map 
between them, both can be naturally simulated in each other. 
Semantically, there is no embedding between the corresponding
institutions.
Along the way, the notions of entailment and satisfaction in 
Meseguer's rewriting logic are generalized.
We also use the syntactic results to prove reflective
properties of CRWL. 
\end{abstract}

\begin{keywords}
rewriting logic, constructor-based rewriting logic, institutions
\end{keywords}

\section{Introduction}

The aim of this paper is to study in detail, and to try to 
clarify, the relationships between two well-known
approaches to rewriting as logical deduction, namely,
Jos\'e Meseguer's rewriting logic \cite{Meseguer92}, 
and the constructor-based
rewriting logic (CRWL) developed by Mario Rodr\'{\i}guez-Artalejo's
research group in Madrid \cite{GonzalezEtAl99}.

The first of these was proposed as a logical framework wherein to
represent other logics, and also as a semantic framework, a unified
model of concurrency for the specification of languages and systems.
The experience accumulated throughout the last years has come to
support that original intention \cite{MartiOlietMeseguer02}. 
In particular, it has been shown that rewriting
logic is a very flexible framework in which many other logics,
including first-order logic, intuitionistic logic, linear logic,
Horn logic with equality, as well as any other logic with a
sequent calculus, can be represented 
\cite{Meseguer00,MartiOlietMeseguer02-rllsf,VerdejoMartiOliet02,%
ThatiEtAl02,Verdejo-thesis}.
An important feature of these representations that should
be stressed is that they are usually quite simple and natural
(in Meseguer's vocabulary, ``the representation distance is zero''),
so that the mathematical properties of the source logics are often 
straightforward to derive in their rewriting logic representation.

On the other hand, the goal of the constructor-based rewriting
logic is to serve as a logical basis for declarative programming
languages involving lazy evaluation, offering support, in addition,
to non-strict and possibly non-deterministic functions.

Despite these differences, there is a clear resemblance between both
logics, namely, the fact that logical deduction is based on 
rewriting.
It seems natural, then, to ask about the relationships between
deduction in these logics and to extend the question so as to
encompass whether the corresponding models are also related.
A suitable framework in which to carry out this study is the theory
of \emph{general logics} developed in \citeN{Meseguer89}.
There, logics are described in a very abstract manner
and two separated components are distinguished:
a syntactic part, which is captured by the notion of 
\emph{entailment system}, and a semantic one, captured by Goguen and
Burstall's concept of \emph{institution} \cite{GoguenBurstall92}.

We will begin by studying derivability and, for that, we will
try to associate entailment systems to both logics and to relate
them by means of a map of entailment systems.
Unfortunately, it will be proved that there is none corresponding
to deduction in CRWL, so we will  be forced to leave this
formal framework and undertake more informal simulations of
the logics in each other.
Although such simulations are always possible by making use of
suitable low-level encodings, relying on the analogies
between both logics our interest resides in finding natural and
simple simulations that at the very least would show that their 
expressive power is the same.
In addition, these results will be used to study reflective properties
of CRWL.

After the comparison at the syntactic level, the next step is
the study of the corresponding models.
Now we will be able to associate an institution to each logic,
so this study will take place within the formal framework of
the theory of institutions.
The main result we will obtain is that models in these logics bear
no relation at all.
Along the way, we generalize the notions of derivability and 
satisfaction in Meseguer's rewriting logic to conditional rewrite
rules, and clarify some subtle points regarding the definition of
models in this logic.

As implied by the previous presentation, this paper does not consider 
the \emph{operational semantics} of the logics, but focus
instead of comparing them at a more abstract level by considering both 
its provability and satisfaction relations. 
We refer to works like \citeN{BoscoEtAl88}, where such operational issues 
related to resolution or narrowing are pursued in similar contexts.

Meseguer's rewriting logic is parameterized with respect to an
underlying equational logic; although this can be typed and as general as
the membership equational logic from \citeN{Meseguer98}, in this paper we 
concentrate on the version of rewriting logic which uses unsorted
and unconditional equational logic and write RL for it.
Likewise, there are also typed versions of CRWL 
\cite{ArenasArtalejo01,GonzalezEtAl01}.
Here we have chosen to focus on the untyped versions because this
work is mainly foundational and the addition of types, while not
introducing any new fundamental concepts, would force us to deal
with many details that would obscure the presentation (for example,
quantifiers would have to be considered explicitly).
The typed cases are considered in some detail 
in \citeN{Palomino-mthesis}.

In what follows we assume familiarity with some basic ideas of category
theory \cite{BarrWells99}, that do not go beyond natural 
transformations and equalizers.
Only in Section~\ref{sec:TMRL} we use a less standard construction,
that of subequalizer, that we describe in the text.

\section{Relations at the Syntactic Level}\label{rsl:sec}

In the first part of the paper we focus on the syntactic aspects
of the logics, and try to abstractly study derivability in them
at the level of entailment systems.
After ruling out this possibility, we develop some simulations
that will allow us to prove some reflective properties of the logics.
We start by reviewing the main concepts and definitions that we
will use.

\subsection{Entailment systems}
\label{sec:ES}

Syntax is typically given by a \emph{signature\/} $\Sigma$ providing
a grammar on which \emph{sentences}, collected in a set
$\Sen(\Sigma)$, are built.
For a given signature $\Sigma$, \emph{entailment\/}
(also called \emph{provability\/}) of a sentence
$\varphi \in \Sen(\Sigma)$ from a set
of axioms $\Gamma \subseteq \Sen(\Sigma)$ is a relation
$\Gamma \vdash \varphi$ which holds if and only if we can prove
$\varphi$ from the axioms $\Gamma$ using the rules of the logic.
We make this relation relative to a signature.
In the rest of the paper, let $|{\cal C}|$ denote the collection of
objects of a category $\cal C$.

An \emph{entailment system} \cite{Meseguer89} is a triple
$\mathcal{E} = (\Sign,\Sen,\vdash)$
such that:
\begin{itemize}
\item
 \Sign\ is a category whose objects are called \emph{signatures}.
\item
 $\Sen: \Sign \to \Set$ is a functor associating
 to each signature $\Sigma$ a corresponding set of 
 $\Sigma$-\emph{sentences}.
\item
 $\vdash$ is a function which associates to each $\Sigma \in |\Sign|$
 a binary relation 
 $\vdash_{\Sigma} \subseteq \mathcal{P}(\Sen(\Sigma)) \times
 \Sen(\Sigma)$ called $\Sigma$-\emph{entailment\/} such that
 the following properties are satisfied: \label{pg:entailment}
 \begin{enumerate}
 \item {\it reflexivity}:
  for any $\varphi \in \Sen(\Sigma)$, $\{\varphi\} \vdash_{\Sigma} \varphi$,
 \item {\it monotonicity}:
  if $\Gamma \vdash_{\Sigma} \varphi$ and
  $\Gamma' \supseteq \Gamma$ then $\Gamma' \vdash_{\Sigma} \varphi$,
 \item {\it transitivity}:
  if $\Gamma \vdash_{\Sigma} \varphi_{i}$, for all $i \in I$, and
  $\Gamma \cup \{\varphi_i \mid i \in I \} \vdash_{\Sigma} \psi$,
  then $\Gamma \vdash_{\Sigma} \psi$,
 \item $\vdash$-{\it translation}:
  if $\Gamma \vdash_{\Sigma} \varphi$, then
  for any $H: \Sigma \to \Sigma'$ in \Sign,
  $\Sen(H)(\Gamma) \vdash_{\Sigma'} \Sen(H)(\varphi)$.
 \end{enumerate}
\end{itemize}

Given an entailment system $\mathcal{E}$, its category \Th\ of
\emph{theories} has as objects pairs $T=(\Sigma,\Gamma)$, with
$\Sigma$ a signature and $\Gamma \subseteq \Sen(\Sigma)$.
A \emph{theory morphism} $H:(\Sigma,\Gamma)\to (\Sigma',\Gamma')$ is
a signature morphism $H:\Sigma\to \Sigma'$ such that if
$\varphi\in \Gamma$, then $\Gamma'\vdash_{\Sigma'} \Sen(H)(\varphi)$.
A theory morphism is \emph{axiom-preserving} if, in addition,
it satisfies the condition $\Sen(H)(\Gamma) \subseteq \Gamma'$.
This defines a subcategory $\Th_0$ with the same objects as \Th\
but with morphisms restricted to be axiom-preserving theory
morphisms, that does not depend on the entailment relation.

Although we will not be able to use it, there is also a notion of 
\emph{map of entailment systems},
allowing us to relate logics in a general and systematic way.
Basically, a map of entailment systems $\mathcal{E} \to \mathcal{E}'$
maps signatures of $\mathcal{E}$  to signatures of $\mathcal{E}'$
(or, more generally, theories to theories),
and sentences of $\mathcal{E}$ to sentences of $\mathcal{E}'$, 
respecting the entailment
relations $\vdash$ of $\mathcal{E}$ and $\vdash'$ of ${\cal E}'$.
See \citeN{Meseguer89} for details.

\subsection{Rewriting logic}

A signature in RL is a pair $(\Sigma,E)$ with $\Sigma$ a ranked
alphabet of function symbols and $E$ a set of $\Sigma$-equations.
Rewriting operates on equivalence classes of terms modulo the
set of equations $E$.
We denote by $T_\Sigma(\mathcal{X})$ the $\Sigma$-algebra of
$\Sigma$-terms with variables in a set $\mathcal{X}$, and by
$[t]_E$ or just $[t]$ the $E$-equivalence class of
$t\in T_\Sigma(\mathcal{X})$.
To indicate that $\{x_1,\dots,x_n\}$ is the set of variables
occurring in $t$ we write $t(x_1,\dots,x_n)$.
Given $t(x_1,\dots,x_n)$, and terms $u_1,\dots,u_n$,
$t(u_1/x_1,\dots,u_n/x_n)$ denotes the term obtained from
$t$ by \emph{simultaneously substituting} $u_i$ for $x_i$,
$i=1,\dots,n$.
To simplify notation we denote a sequence of objects $a_1,\dots,a_n$
by $\overline{a}$; with this notation, $t(u_1/x_1,\dots,u_n/x_n)$
can be abbreviated to $t(\overline{u}/\overline{x})$. 

An RL-theory $\mathcal{R}$ is a 4-tuple
$\mathcal{R} = (\Sigma,E,L,\Gamma)$, where $(\Sigma,E)$ is a
signature and $\Gamma$ is a set of rewrite rules, \emph{labeled}
with elements of $L$, of the form
\begin{displaymath}
r: [t]\to [t'] \ifcond [a_1] \to [b_1] \land \dots \land 
                       [a_m] \to [b_m] \,\textrm{.}
\end{displaymath}
We write $\mathcal{R} \vdash [t] \to [t']$ if $[t]\to [t']$ can
be derived using the rules of deduction shown 
in Figure~\ref{fig:calculi1};
for a complete exposition of RL we refer the reader 
to \citeN{Meseguer92}.
\begin{figure}
\figrule
\begin{displaymath}
\begin{array}{c}
\bigfrac{}{[t] \to [t]}\; \textbf{Reflexivity}\qquad
\bigfrac{[t]\to [t']\quad [t']\to [t'']}{[t]\to [t'']} \;
\textbf{Transitivity} \\
\noalign{\bigskip}     
\bigfrac{[t_1]\to [t_1'] \ \dots\ [t_n]\to [t_n']}
     {[f(t_1,\dots,t_n)]\to [f(t_1',\dots,t_n')]} \;
\textbf{Congruence} \\
\noalign{\bigskip}     
\bigfrac{\bigatop{r : [t] \to [t'] \ifcond
                      [a_1]\to [b_1] \land \dots\land
                      [a_m]\to [b_m] \in \Gamma}
                 {{\begin{array}{rcl}
                   [w_1]\to [w_1'] &\dots &[w_n]\to [w_n'] \\
                   {[}a_1(\overline{w}/\overline{x})]\to 
                       [b_1(\overline{w}/\overline{x})] &\dots
                     & [a_m(\overline{w}/\overline{x})]\to
                       [b_m(\overline{w}/\overline{x})] 
                   \end{array}}}}
        {{[t(\overline{w}/\overline{x})]\to 
         [t'(\overline{w'}/\overline{x})]}}\; \textbf{Replacement}
\end{array}
\end{displaymath}
\caption{Rules of deduction for an RL-theory $(\Sigma,E,L,\Gamma)$}
\figrule
\label{fig:calculi1}
\end{figure}

\subsection{CRWL}
\label{sec:CRWL}

CRWL uses signatures with constructors
$\Sigma = C_\Sigma \cup F_\Sigma$, where 
$C_\Sigma = \bigcup_{n\in \nat} C_\Sigma^n$ and
$F_\Sigma = \bigcup_{n\in \nat} F_\Sigma^n$
are disjoint sets of \emph{constructor} and \emph{defined function
symbols} respectively, each of them with an associated arity.
$\Sigma_\bot$ refers to the signature which is obtained from
$\Sigma$ by adding a new constructor $\bot$ of arity 0.
Given a set $\mathcal{X}$ of variables, we will write
\Expr{\Sigma}{X} for the set of total expressions which can be
built with $\Sigma$ and $\mathcal{X}$, and \Term{\Sigma}{X}
for those total terms which only make use of $C_\Sigma$ and $\mathcal{X}$.
\Bexpr{\Sigma}{X} and \Bterm{\Sigma}{X}, the sets of \emph{partial}
expressions and terms, are defined analogously using $\Sigma_\bot$.
A \emph{signature morphism} \cite{Molina-thesis}
$\sigma:\Sigma\to\Sigma'$ from a
signature $\Sigma=C_\Sigma \cup F_\Sigma$ to another
$\Sigma'=C_{\Sigma'} \cup F_{\Sigma'}$ is a pair of functions
(denoted with the same $\sigma$)
\begin{displaymath}
\sigma:C_\Sigma\to C_{\Sigma'}\quad\textrm{and}\quad
\sigma:F_\Sigma\to F_{\Sigma'},
\end{displaymath}
mapping $n$-ary symbols to $n$-ary symbols.

A CRWL-theory is a pair $(\Sigma,\Gamma)$, where $\Sigma$ is a
signature with constructors and $\Gamma$ is a set of conditional
rewrite rules of the form
\begin{displaymath}
f(t_1,\dots,t_n) \to r \Leftarrow a_1\bowtie b_1,\dots,
                                  a_m\bowtie b_m \quad (m\geq 0),
\end{displaymath}
with $f\in F_\Sigma$ of arity $n$, $t_1,\dots,t_n\in \Term{\Sigma}{X}$,
$r, a_i, b_i \in \Expr{\Sigma}{X}, i=1,\dots,m$, $r$
and each variable
occurring in $t_1,\dots,t_n$ having a single ocurrence.

From a given theory $T$, two kinds of sentences can be derived using
the CRWL-calculus in Figure~\ref{fig:calculi2} (where variables range over
partial expressions): reduction
statements of the form $a\to b$, and joinability statements
$a\bowtie b$ (meaning that there exists a total term to which both
$a$ and $b$ reduce). 
We denote them by $T\vdash a \to b$ and $T\vdash a \bowtie b$, respectively.
Again, we refer to \citeN{GonzalezEtAl99} for a complete presentation 
of CRWL.
(Note that the names ``term'' and ``constructor term'' are used there
instead of ``expression'' and ``term.'')
\begin{figure}
\figrule
\begin{displaymath}
\begin{array}{c}
\bigfrac{}{e\to \bot} \;\textbf{Bottom} \qquad
\bigfrac{}{e\to e} \; \textbf{Reflexivity} \\
\noalign{\bigskip}
\bigfrac{e_1\to e_1'\ \dots\ e_n\to e_n'} 
        {h(e_1,\dots,e_n)\to h(e_1',\dots,e_n')} \; 
        \textbf{Monotonicity}\\
\noalign{\bigskip}     
\bigfrac{
  \bigatop{\theta :\mathcal{X}\longrightarrow\Bterm{\Sigma}{X}}
          {\bigatop{l \to r \Leftarrow a_1 \bowtie b_1,\dots,
                            a_n\bowtie b_n \in \Gamma}
                   {\theta(a_1)\bowtie \theta(b_1)\ \dots\
                     \theta(a_n)\bowtie \theta(b_n)}}}
        {\theta(l)\to \theta(r)} \;
 \textbf{Reduction} \\
\noalign{\bigskip}
\bigfrac{e\to e' \quad e'\to e''}{e\to e''} \; 
 \textbf{Transitivity}\qquad
\bigfrac{a\to t \quad b\to t \quad \textrm{$t$ a total term}}
        {a\bowtie b} \;
 \textbf{Join} 
\end{array}
\end{displaymath}
\caption{Rules of deduction for a CRWL-theory $(\Sigma,\Gamma)$}
\figrule
\label{fig:calculi2}
\end{figure}

\subsection{An entailment system for RL}
\label{sec:ESRLCRWL}

In order to associate an entailment system to RL, note that the rules
of inference in Figure~\ref{fig:calculi1} only allow us to derive
unconditional rules but that the requirements on an entailment system
(\textit{reflexivity}) require the ability to derive conditional ones
as well. 
We then have two possibilities:
either we restrict ourselves to unconditional rewrite rules and
define $\vdash_\Sigma$ by means of derivation in the 
RL-calculus, or we also consider conditional rules, in which case the
RL-calculus in Figure~\ref{fig:calculi1} must be extended 
to be able to derive them.
We consider the second, more general case.

Actually, not only is derivability undefined for conditional
rules, but also is satisfaction.
However, we would like to rest on a natural definition of
satisfaction to support the claim that our extended notion of 
derivability is a suitable one.
The semantics of RL is presented in Section~\ref{sec:TMRL} and the
extension of the satisfaction relation  discussed in 
Section~\ref{sec:IRL}; here we just assume that such an extension 
exists.

Given an RL-theory $\mathcal{R} = (\Sigma,E,L,\Gamma)$ and
a set of variables $\mathcal{X}$ disjoint from $\Sigma$, we define
$\mathcal{R}(\mathcal{X}) = (\Sigma(\mathcal{X}),E,L,\Gamma')$ where 
$\Sigma(\mathcal{X})$
is the set of function symbols obtained by adding the elements
of $\mathcal{X}$ as constants to $\Sigma$, and $\Gamma'$ is obtained from
$\Gamma$ by renaming with fresh variables.
In Section~\ref{sec:IRL} it is proved that,
for an RL-theory $\mathcal{R}$ and
$[t(\overline{x})]\to [t'(\overline{x})] \ifcond
 [a_1(\overline{x})] \to [b_1(\overline{x})]\land \dots \land
 [a_m(\overline{x})] \to [b_m(\overline{x})]$ a conditional rewrite 
rule, the following statements are equivalent:
\begin{enumerate}
\item $\mathcal{R} \models [t(\overline{x})]\to [t'(\overline{x})]
 \ifcond
 [a_1(\overline{x})] \to [b_1(\overline{x})]\land \dots \land
 [a_m(\overline{x})] \to [b_m(\overline{x})]$;
\item $\mathcal{R}(\overline{x}) \cup
 \{ [a_1(\overline{x})] \to [b_1(\overline{x})], \dots,
    [a_m(\overline{x})] \to [b_m(\overline{x})] \}
  \models [t(\overline{x})] \to [t'(\overline{x})]$.
\end{enumerate}

A straightforward consequence of this equivalence
is a sound and complete extension of the RL-calculus
with the following rule of deduction:
\begin{itemize}
\item \textbf{Implication introduction}.
\begin{displaymath}
\bigfrac{\mathcal{R}(\overline{x}) \cup
         \{ [a_1(\overline{x})]\to [b_1(\overline{x})], \dots,
            [a_m(\overline{x})]\to [b_m(\overline{x})] \}
            \vdash
            [t(\overline{x})]\to [t'(\overline{x})]}
 {\mathcal{R} \vdash
    [t(\overline{x})]\to [t'(\overline{x})] \ifcond
    [a_1(\overline{x})]\to [b_1(\overline{x})] \land\dots \land
    [a_m(\overline{x})]\to [b_m(\overline{x})]}.
\end{displaymath}
\end{itemize}

We can now focus again on the main purpose of this section.
For that, we associate to RL the entailment sytem
$\mathcal{E}_\mathrm{RL} = (\Sign, \Sen, \vdash)$
given by:
\begin{itemize}
\item \Sign: the category of equational theories and theory
 morphisms;
\item \Sen: the functor assigning to an equational theory the
 set of conditional rewrite rules that can be built over it, and
 mapping a theory morphism to its natural extension to rewrite rules;
\item $\vdash$ is defined as provability in the extended RL-calculus.
\end{itemize}
\begin{proposition}
$\mathcal{E}_\mathrm{RL} = (\Sign, \Sen, \vdash)$ is
an entailment system.
\end{proposition}
The proof of this result uses concepts from the model theory of RL
that are not introduced until later in the text, so we postpone the
details to the appendix.

Before finishing,
it should be emphasized that throughout this section no mention at all
has been made of the labels in an RL-theory.
They could have been safely included within the signature part;
however, they do not play any role as far as the entailment relation
is concerned and, if only for ease of exposition, we have preferred
to omit then.
This situation will change drastically when we shift to models
and try to assign an institution to RL; then, we will be forced
to distinguish between labeled and unlabeled sentences, as
described in Section~\ref{sec:IRL}.

\subsection{An entailment system for CRWL}
\label{sec:AESCRWL}

At first sight, an entailment system can be associated to CRWL 
following the same steps as for RL.
The category of signatures is immediately obtained, as it is
not difficult to check that composition of signature
morphisms is associative, and for the set of sentences we have
the same two possibilities as for RL.

However, a closer look reveals that derivation in the 
CRWL-calculus does not satisfy the transitivity condition for the
provability relation in entailment systems.
Consider, for example, a signature $\Sigma$ with $c,d,h\in \Sigma$,
function symbols of arities 0, 0, and 1, respectively.
Then it can be proved that
\begin{displaymath}
\{\, c \to h(c), h(x) \to h(d) \Leftarrow x \bowtie x
\,\} \vdash_\mathrm{CRWL} h(x) \to h(d)
\end{displaymath}
and
\begin{displaymath}
\{\,c\to h(c), h(x) \to h(d)\Leftarrow x\bowtie x, h(x)\to h(d) \,\}
 \vdash_\mathrm{CRWL} c \to h(d),
\end{displaymath}
but
\begin{displaymath}
\{\, c \to h(c), h(x) \to h(d) \Leftarrow x \bowtie x \,\}
\not\vdash_\mathrm{CRWL} c \to h(d)\,\textrm{.}
\end{displaymath}
The first statement is proved by instantiating 
$h(x)\to h(d) \Leftarrow x\bowtie x$ with $x$;
for the second, noting that $h(c) \to h(\bot)$ (using \textbf{Bottom}
and \textbf{Congruence}), just instantiate $h(x)\to h(d)$ 
with $\bot$ and apply \textbf{Transitivity} (note that $c$ cannot 
be used to instantiate this rule since it is not a term).
The third statement is formally proved by induction on derivations:
let us just note that the crucial point is that the rule
$h(x)\to h(d) \Leftarrow x \bowtie x$ cannot be instantiated with $\bot$
because $\bot \bowtie \bot$ cannot be derived.
What lies behind is the fact 
that the CRWL-calculus is sound and complete
with respect to validity in models \emph{only} under totally defined
valuations \cite{GonzalezEtAl99}.
In particular, in the first entailment above, $h(x)\to h(d)$
means that $h(t)$ rewrites to $h(d)$ just for those instances
where a \emph{total} term $t$ is substituted for $x$.

This proves that the relation $\vdash_\mathrm{CRWL}$ is not
transitive and, therefore, we are not going to be able to build an
entailment system \emph{based on} the CRWL-calculus, as
any sensible one should contain, at least, the conditional 
rewrite rules among its sentences. 
(Let us note, however, that there is an entailment system 
corresponding to the institution that will be associated to CRWL in 
Section~\ref{sec:ICRWL}; 
the previous example is no longer a counterexample due to the
partiality of the soundness and completeness results for CRWL 
mentioned above.
The reason for not comparing it to RL's entailment system is that,
since it is not based on deduction, they do not stand ``on the same 
ground.'')

\subsection{Simulating CRWL in RL}
\label{sec:SCRWLRL}

Since there is no entailment system corresponding to the
CRWL-calculus, we cannot define a map of entailment systems as
intended.
In the following we will be pleased just with presenting how
entailment in CRWL can be simulated in RL.
The set of labels of an RL-theory does not take part
in the entailment process, and so it is omitted; the same convention 
will also be adopted in Section~\ref{sec:SRLCRWL}.

Of course, every CRWL-theory $T$ can be trivially ``simulated'' in RL
by means of an RL-theory $T'$ with a constant $c_t$ for each term
(and each expression) $t$ in $T$, and with axioms $c_t\to c_{t'}$
whenever $T\vdash t\to t'$.
But such a $T'$, apart from not exploiting the analogies 
between RL and CRWL, is not computable in general.
And so we must look for another construction.

The idea is to associate to every CRWL-theory $T=(\Sigma,\Gamma)$ 
a theory $T'$ in
RL (whose set of equational axioms will be empty)
in which all the operations in $T$, together with a new
constant $\bot$, are available, plus one rule for each axiom in $T$
and, perhaps, some more rules coping with the rules of deduction
of the CRWL-calculus.
Since rules in CRWL can only be instantiated with terms and
not expressions and there is no such distinction in RL, we introduce
a unary relation \pterm\
(technically, a unary function symbol) and a constant \True\ to 
distinguish them in RL.
One immediate rule defining \pterm\ is $\pterm(\bot)\to\True$;
however, how to express that variables are also partial terms?
The obvious rule $\pterm(x)\to \True$ is clearly not valid: 
everything would be a partial term!
This means that we must consider the CRWL variables at the object
level, add them to the signature
of $T'$ as constants, and use a new set $\mathcal{X}$ of variables
for RL.
Using constants for variables we will be able to distinguish those
terms in RL representing terms in CRWL from those representing
expressions, hence allowing us to capture, by carefully 
translating the rules of deduction of the CRWL-calculus (using,
perhaps, a different representation for the terms appearing in
them), the corresponding entailment relation.

Then, assuming variables in CRWL belong to a set $\mathcal{V}$,
the rules defining \pterm\ are:
\begin{displaymath}
\begin{array}{l}
 \pterm(\bot)\to\True \\
 \pterm(v_i)\to\True\quad(\forall v_i\in\mathcal{V}) \\
 \pterm(h(x_1,\dots,x_n))\to\True \\
 \qquad\ifcond \pterm(x_1)\to\True \land\dots\land\pterm(x_n)\to\True\quad
 (\forall h\in C_\Sigma^n, n\in\nat)
\end{array}
\end{displaymath}
In a similar way, two more predicates, \tterm\ and \pexpr, dealing
with total terms and partial expressions, are defined:
\begin{displaymath}
\begin{array}{l}
 \tterm(v_i)\to\True \quad(\forall v_i\in\mathcal{V}) \\
 \tterm(h(x_1,\dots,x_n))\to\True \\
 \qquad \ifcond \tterm(x_1)\to\True \land\dots\land
 \tterm(x_n)\to\True\quad
       (\forall h\in C_\Sigma^n, n\in\nat) \\
 \pexpr(\bot)\to\True \\
 \pexpr(v_i)\to\True\quad(\forall v_i\in\mathcal{V}) \\
 \pexpr(h(x_1,\dots,x_n))\to\True \\
 \qquad\ifcond \pexpr(x_1)\to\True \land\dots\land
               \pexpr(x_n)\to\True\quad
       (\forall h\in C_\Sigma^n\cup F_\Sigma^n, n\in \nat) \\
\end{array}
\end{displaymath}

As a side effect, rewriting in CRWL can no longer be simulated 
in RL directly through the rewriting relation.
Consider, for example, the theory of natural numbers in CRWL,
with 0 a constructor and $+$ a function symbol.
In RL, $\pterm(0+0)$ should not rewrite to \True; however, with
the usual definitions, $0+0\to 0$ and by \textbf{Congruence}
$\pterm(0+0)\to \pterm(0)$, and this last term must reduce to \True.
Therefore,  a rewrite in CRWL will be simulated through a binary 
relation $R$ so that $e\to e'$ in CRWL if and only if 
$R(e,e')\to\True$ in RL.
In a similar way, strict equalities $a\bowtie b$ will be 
simulated through a binary relation $\bowtie$.

It just remains to translate the rules of deduction of the CRWL-calculus,
which is straightforward.
For example, the \textbf{Bottom} rule stating that every expression is 
reducible to $\bot$ is written
\begin{displaymath}
R(x,\bot) \to \True \ifcond \pexpr(x) \to \True,
\end{displaymath}
whereas the \textbf{Join} rule
\begin{displaymath}
\bigfrac{a\to t \qquad b\to t \qquad\textrm{$t$ is a total term}}
        {a\bowtie b} 
\end{displaymath}
results in
\begin{displaymath}
 x\bowtie y\to\True
 \ifcond R(x,z)\to\True\land R(y,z)\to\True\land \tterm(z)\to\True\,
 \textrm{.}
\end{displaymath}
\textbf{Reflexivity}, \textbf{Transitivity}, and \textbf{Monotonicity}
are taken care of by, respectively:
\begin{displaymath}
\begin{array}{l}%
R(x,x)\to\True \ifcond \pexpr(x)\to\True \\
R(x,y)\to\True \ifcond R(x,z)\to\True\land R(z,y)\to\True\\
R(h(x_1,\dots,x_n),h(y_1,\dots,y_n))\to\True\\
\qquad \ifcond R(x_1,y_1)\to\True \land\dots\land 
               R(x_n,y_n)\to\True\quad 
       (\forall h\in C_\Sigma^n\cup F_\Sigma^n, n\in\nat)
\end{array}
\end{displaymath}
It is not necessary to include \pexpr\ explicitly in
all the rules because these conditions can be derived as
logical consequences.

Finally, to every rule $l(\overline{v}) \to r(\overline{v}) \Leftarrow
a_1(\overline{v}) \bowtie b_1(\overline{v}), \dots,
a_m(\overline{v}) \bowtie b_m(\overline{v})$ in the CRWL-theory,
we associate the following rule in RL,
\begin{displaymath}
\begin{array}{l}
 R(l(\overline{x}),r(\overline{x}))\to\True\\
 \qquad\ifcond a_1(\overline{x})\bowtie b_1(\overline{x})\to\True
               \land\dots\land 
               a_m(\overline{x})\bowtie b_m(\overline{x})\to\True\land{}\\
 \qquad \phantom{\ifcond}
        \pterm(x_1)\to\True \land\dots\land 
        \pterm(x_n)\to\True\,,
\end{array}
\end{displaymath}
where each CRWL variable $v_i$ (a constant in the RL-theory) has been
replaced by the variable $x_i$.
The set of all these last rules corresponds to the \textbf{Reduction}
rule in the CRWL-calculus,
and the condition that program rules in CRWL can
only be instantiated with terms is taken care of by
demanding $\pterm(x)\to\True$ for all the variables appearing
in it.

We will write $\alpha(T)= (\Sigma', \emptyset, \Gamma')$ for the RL-theory 
associated to a CRWL-theory $T$ in this fashion, with
$\Sigma' =\Sigma\cup \mathcal{V} \cup \{\True,\pterm,\pexpr,\tterm, 
R,\bowtie,\bot\}$ 
and $\Gamma'$ consisting of all the rules described above. 
The following proposition ensures that the translation
is correct.
Note that we use $=$ to denote both syntactic and semantic equality: the 
context will always make clear to which one we refer.
\begin{proposition}
\label{prop:CRWLinRLentailment}
For a CRWL-theory $T=(\Sigma,\Gamma)$ with
$\alpha(T)=(\Sigma',\emptyset,\Gamma')$, 
if $l, r, a, b\in T_{\Sigma'}(\mathcal{X})$:
\begin{displaymath}
\begin{array}{rcl}
l,r\in\Bexpr{\Sigma}{V} \textrm{ and } T\vdash_\mathrm{CRWL} l\to r
 &\iff &\alpha(T)\vdash_\mathrm{RL} R(l,r)\to\True\,; \\
a,b\in\Bexpr{\Sigma}{V} \textrm{ and } T\vdash_\mathrm{CRWL} a\bowtie b
	 &\iff &\alpha(T)\vdash_\mathrm{RL} a\bowtie b\to\True\,\textrm{.}
\end{array}
\end{displaymath}
\end{proposition}

The following two lemmas, which can be easily proved by structural induction 
on derivations, are needed for its proof.
In particular, Lemma~\ref{lem:SCRWLRLaux1} is used in the most difficult part, 
which corresponds to \textbf{Transitivity} in the
$(\Leftarrow)$-direction.
\begin{lemma}
\label{lem:SCRWLRLaux1}
Let $T=(\Sigma,\Gamma)$ be a CRWL-theory, $\alpha(T)=(\Sigma',\emptyset,
\Gamma')$,
and $e,e'\in T_{\Sigma'}(\mathcal{X})$.
\begin{enumerate}
\item If\/ $\alpha(T)\vdash_\mathrm{RL} e\to e'$ and $e\in\Bexpr{\Sigma}{V}$
 or $e'\in\Bexpr{\Sigma}{V}$, then $e = e'$.
\item If\/ $\alpha(T)\vdash_\mathrm{RL} \tterm(e)\to e'$, then 
 $e'$ is either \True\ or $\tterm(e'')$ for some $e''$ such
 that $\alpha(T)\vdash_\mathrm{RL} e\to e''$. 
\end{enumerate}
\end{lemma}

\begin{lemma}
\label{lem:SCRWLRLaux2}
If $T=(\Sigma,\Gamma)$ is a CRWL-theory, $\alpha(T)=(\Sigma',\emptyset,
\Gamma')$, and $e\in T_{\Sigma'}(\mathcal{X})$, then:
\begin{enumerate}
\item $e\in\Term{\Sigma}{V} \iff
 \alpha(T)\vdash_\mathrm{RL}\tterm(e)\to\True$,
\item $e\in\Bterm{\Sigma}{V} \iff
 \alpha(T)\vdash_\mathrm{RL}\pterm(e)\to\True$,
\item $e\in\Bexpr{\Sigma}{V}
 \iff \alpha(T)\vdash_\mathrm{RL}\pexpr(e)\to\True$.
\end{enumerate}
\end{lemma}

\begin{proof*}[Proof of Proposition~\ref{prop:CRWLinRLentailment}]
Both directions are proved by induction on the derivation, studying
the last rule applied.
Let us first consider the ($\Rightarrow$) part.
\begin{itemize}
\item \textbf{Bottom}. We have $T\vdash_\mathrm{CRWL} l\to\bot$.
 Since $l\in \Bexpr{\Sigma}{V}$, by Lemma~\ref{lem:SCRWLRLaux2} 
 it is $\alpha(T)\vdash_\mathrm{RL} \pexpr(l)\to\True$
 so, by the translation of the \textbf{Bottom} rule, we have
 $\alpha(T)\vdash_\mathrm{RL} R(l,\bot)\to\True$.
\item \textbf{Reflexivity}. By Lemma~\ref{lem:SCRWLRLaux2},
 $\alpha(T)\vdash_\mathrm{RL} \pexpr(l)\to\True$, so the result
 follows by applying the third rule associated to the relation $R$.
\item \textbf{Transitivity}. We have that the last step in the 
 derivation is
 \begin{displaymath}
 \bigfrac{a\to t\qquad t\to r}
         {l\to r}.
 \end{displaymath}
 By induction hypothesis, $\alpha(T)\vdash_\mathrm{RL} R(l,t)\to\True$ and
 $\alpha(T)\vdash_\mathrm{RL} R(t,r)\to\True$, and by
 the fourth rule associated to $R$ we can derive
 $\alpha(T)\vdash_\mathrm{RL} R(l,r)\to\True$.
\item \textbf{Monotonicity}. Similarly to the previous case.
\item \textbf{Join}. From
 \begin{displaymath}
 \bigfrac{a\to t \qquad b\to t}
         {a\bowtie b} \quad t\in\Term{\Sigma}{V},
 \end{displaymath}
 we get, by induction hypothesis,
 $\alpha(T)\vdash_\mathrm{RL} R(a,t)\to\True$ and
 $\alpha(T)\vdash_\mathrm{RL} R(b,t)\to\True$, and by 
 Lemma~\ref{lem:SCRWLRLaux2},
 $\alpha(T)\vdash_\mathrm{RL} \tterm(t)\to\True$,
 so we can apply the
 rule associated to $\bowtie$ to reach the result.
\item \textbf{Reduction}. Assume that for some rule 
 $l(\overline{v}) \to r(\overline{v}) \Leftarrow
  a_1(\overline{v}) \bowtie b_1(\overline{v}), \dots,
  a_m(\overline{v}) \bowtie b_m(\overline{v})$ in $\Gamma$  and 
 partial terms $w_1,\dots,w_n$, the last step in the derivation is
 \begin{displaymath} 
 \bigfrac{a_1(\overline{w}/\overline{v})\bowtie 
          b_1(\overline{w}/\overline{v}) \;\;\dots\;\;
          a_m(\overline{w}/\overline{v})\bowtie 
          b_m(\overline{w}/\overline{v})}
         {l(\overline{w}/\overline{v}) \to 
          r(\overline{w}/\overline{v})}.
 \end{displaymath}
 Then, by induction hypothesis, 
 $\alpha(T)\vdash a_i(\overline{w}/\overline{v})\bowtie 
                  b_i(\overline{w}/\overline{v}) \to \True$ for
 $i = 1,\dots, m$ and, by Lemma~\ref{lem:SCRWLRLaux2}, 
 $\alpha(T) \vdash\pexpr(w_i) \to \True$ for $i=1,\dots,n$.
 Then the result follows by using the rule associated to 
 $l(\overline{v}) \to r(\overline{v}) \Leftarrow
  a_1(\overline{v}) \bowtie b_1(\overline{v}), \dots,
  a_m(\overline{v}) \bowtie b_m(\overline{v})$
 in $\alpha(T)$.
\end{itemize}
Let us now consider the converse ($\Leftarrow$).
Note that even though the names of some of the rules are the
same, the induction proceeds now over derivations in RL.
\begin{itemize}
\item \textbf{Reflexivity} and \textbf{Congruence} are not possible.
\item \textbf{Transitivity}. Assume that 
 \begin{displaymath}
 \bigfrac{R(l,r)\to e\qquad  e\to \True}
         {R(l,r)\to\True}.
 \end{displaymath}
 (The case for $a\bowtie b$ is analogous.)
 By induction on the derivation of
 $\alpha(T)\vdash_\mathrm{RL} R(l,r)\to e$ and using the fact
 that \True\ only rewrites to itself, it follows easily
 that $e$ must be either \True,
 or $R(l',r')$ with $\alpha(T)\vdash_\mathrm{RL} l\to l'$ and
 $\alpha(T)\vdash_\mathrm{RL} r\to r'$.
 In the first case the result follows from the induction 
 hypothesis applied to $R(l,r) \to e$.
 In the second, again by the
 induction hypothesis, $l',r'\in\Bexpr{\Sigma}{V}$ and
 $T\vdash_\mathrm{CRWL} l'\to r'$, and by
 Lemma~\ref{lem:SCRWLRLaux1} we have $l=l'$ and $r=r'$.
\item \textbf{Replacement}. The result follows because the rules 
 associated to the relation 
 $R$ reflect faithfully the rules of deduction of CRWL.
 For example, consider the rule associated to \textbf{Join}.
 If
 \begin{displaymath}
 \bigfrac{R(a,c)\to\True\qquad R(b,c)\to\True \qquad\tterm(c)\to\True}
         {a \bowtie b\to\True},
 \end{displaymath}
 then $c\in \Term{\Sigma}{V}$ by Lemma~\ref{lem:SCRWLRLaux2}, and
 $T\vdash_\mathrm{CRWL} a\to c$ and $T\vdash_\mathrm{CRWL} b\to c$
 by the induction hypothesis, whence follows that 
 $T\vdash_\mathrm{CRWL} a \bowtie b$.$\mathproofbox$
\end{itemize}
\end{proof*}

\subsection{Simulating RL in CRWL}
\label{sec:SRLCRWL}

We now embark ourselves on finding the converse simulation of
RL in CRWL.
We are again interested in a computable and simple translation, and
the idea for this is very similar to that of the previous section.
Now, however, there are no terms and expressions to distinguish, 
and therefore predicates such as \pterm\ are no longer necessary; 
as a consequence, we will be able to use the same set $\mathcal{X}$
of variables for both logics.
The fact that only joinability statements are allowed to appear in
the condition of a rewrite rule in CRWL forces us to represent,
as in Section~\ref{sec:SCRWLRL},
the rewriting relation in RL through a binary relation $R$
in CRWL, so that $t\to t'$ in RL if and only
if $R(t,t')\to\True$ in CRWL.
Rewriting modulo a set of equations
will be handled by transforming each equation $t=t'$
into the rewrites $t\to t'$ and $t'\to t$.

More precisely, given a signature $(\Sigma,E)$ in RL we associate
to it a CRWL-theory over the signature $\Sigma'$ with
$C_{\Sigma'} = \Sigma\cup\{\True\}$ and $F_{\Sigma'}=\{R\}$,
with \True\ and $R$ of arities 0 and 2, respectively.
The rules in the theory include 
\begin{displaymath}
\begin{array}{l}
R(x_1,x_2) \to \True \Leftarrow x_1 \bowtie x_2\,, \\
R(x,y) \to \True     \Leftarrow R(x,z)\bowtie\True, R(z,y)\bowtie\True\,,
\end{array}
\end{displaymath}
and, for each $f\in \Sigma$ of arity $n\in \nat$,
\begin{displaymath}
\begin{array}{l}
 R(f(x_1,\dots,x_n),f(y_1,\dots,y_n)) \to \True \\
 \qquad\qquad\qquad \Leftarrow R(x_1,y_1)\bowtie \True,\dots, R(x_n,y_n)\bowtie\True,
\end{array}
\end{displaymath}
mimicking the \textbf{Reflexivity}, \textbf{Transitivity}, and 
\textbf{Congruence} rules in the RL-calculus, together with
\begin{displaymath}
\begin{array}{l}
R(t,t') \to \True\,, \\
R(t',t) \to \True\,,
\end{array}
\end{displaymath}
for every $t=t' \in E$.
The goal of the condition in the rule corresponding to 
\textbf{Reflexivity} is
to avoid instantiating it with terms containing $\bot$, which have no
meaning in RL.

A conditional rewrite rule
\begin{displaymath}
[l] \to [r] \ifcond [a_1]\to[b_1]\land \dots\land [a_m]\to [b_m]
\end{displaymath}
over $(\Sigma,E)$ in RL is then translated to
\begin{displaymath}
R(l,r)\to\True \Leftarrow R(a_1,b_1)\bowtie \True,\dots,
                          R(a_m,b_m)\bowtie \True\,,
\end{displaymath}
where $l$, $r$, $a_i$, $b_i$ are arbitrary members of
$[l]$, $[r]$, $[a_i]$, and $[b_i]$, respectively.
Then, given an RL-theory $T = (\Sigma,E,\Gamma)$ we associate to
it the CRWL-theory $\beta(T)$ obtained by adding to the 
CRWL-theory corresponding to $(\Sigma,E)$ the translation
of the rules in $\Gamma$.

Actually, the previous definitions must be slightly modified
due to some technical details.
Recall from Section~\ref{sec:CRWL} that
in a conditional rewrite rule $l\to r \Leftarrow C$ in CRWL
$l$ must be linear, and it is obvious that with the above definitions
this property is not ensured for the translation of equations
and rewrite rules;
therefore, those rules must be ``linearised''
\cite{ArenasArtalejo01}.
The linearised version of a conditional rewrite rule
$l\to r \Leftarrow C$ is given by
$l'\to r \Leftarrow C,C_l$, where $l'$ and $C_l$ are calculated
as follows: for every variable $x$ appearing $k>1$ times in $l$,
its $j$-th occurrence, $2\leq j\leq k$, is replaced with
a new variable $y_j$ and
$x\bowtie y_j$ is added to $C_l$.
Moreover (and this is simply a feature of our translation),
even when a variable $x$ appears only once, $x\bowtie x$ will be
added to the conditional part so that $x$ cannot be instantiated
with a partial term.
The treatment of linearised rules in the rest of the section,
though rigorous, will not be too formal.

The following proposition shows that our translation correctly
reflects provability in the equational part of a rewrite theory.

\begin{proposition} \label{prop:aux}
If $(\Sigma',E')$ is the CRWL theory  corresponding to a
signature $(\Sigma,E)$ in RL and if 
$t,t'\in T_\Sigma(\mathcal{X})$ are such that $E \vdash t=t'$, then
\begin{displaymath}
 E'\vdash_\mathrm{CRWL} R(t,t')\to\True \qquad\textrm{and}\qquad
 E'\vdash_\mathrm{CRWL} R(t',t)\to\True\,\textrm{.} 
\end{displaymath}
\end{proposition}
\begin{proof}
By induction on the derivation of $E\vdash t=t'$.
The rules of a deduction system for equational logic 
include those in Figure~\ref{fig:calculi1} (replacing the arrow
with an equality symbol) together with a \textbf{Symmetry} rule.
Let us just consider the case of the \textbf{Replacement} rule.
Since our equational logic is unconditional, we have
\begin{displaymath}
\bigfrac{}{\theta(t_1) = \theta(t_2)} \qquad (t_1 = t_2) \in E,
\end{displaymath}
for some assignment $\theta: \mathcal{X} \to T_\Sigma(\mathcal{X})$.
Associated to $t_1=t_2$ we have the linearised versions of the two rules
$R(t_1,t_2)\to \True$ and $R(t_2,t_1)\to \True$ in $E'$ and, since
$T_\Sigma(\mathcal{X}) \subseteq \Bterm{\Sigma'}{X}$, we can
instantiate them with $\theta$ (mapping those $x$ which arose in the
linearization process to the same term as the original variable)
to obtain the result.
\end{proof}

With this in hand we are ready to prove the first half of the main 
proposition, which guarantees the correctness of the translation.
\begin{proposition}
\label{prop:RLinCRWLhalf}
Given any RL-theory $T=(\Sigma,E,\Gamma)$, and
$l,r\in T_\Sigma(\mathcal{X})$:
\begin{displaymath}
\begin{array}{rcl}
T\vdash_\mathrm{RL} [l]\to[r] &\Rightarrow
 &(\exists l'\in[l],\exists r'\in [r]) \; \;
 \beta(T)\vdash_\mathrm{CRWL} R(l',r')\to\True  \\
&\iff &(\forall l'\in[l],\forall r'\in [r]) \; \;
 \beta(T)\vdash_\mathrm{CRWL} R(l',r')\to\True
\end{array}
\end{displaymath}
\end{proposition}
\begin{proof*}
Let us first prove the equivalence.
There is nothing to prove in the right-to-left direction; in the 
opposite one, the result is a consequence of 
Proposition~\ref{prop:aux} and the rule
$R(x,y) \to \True \Leftarrow R(x,z)\bowtie\True, R(z,y)\bowtie\True$ 
that, by construction, is in $\beta(T)$.
Now we prove the first implication by induction on the derivation,
according to the last rule used:
\begin{itemize}
\item \textbf{Reflexivity}. 
 $T\vdash_\mathrm{RL} [l]\to [l]$, and the result follows
 by instantiating $R(x_1,x_2)\to\True\Leftarrow x_1\bowtie x_2$
 with $l$ for both variables.
\item \textbf{Congruence}. From
 \begin{displaymath}
 \bigfrac{[l_1]\to [r_1] \;\;\; \dots \;\;\; [l_n]\to [r_n]}
         {[f(l_1,\dots,l_n)]\to [f(r_1,\dots,r_n)]}
 \end{displaymath}
 and the induction hypothesis,
 $\beta(T)\vdash_\mathrm{CRWL} R(l_i',r_i')\to\True$
 for some $l_i'\in [l_i]$, $r_i'\in [r_i]$, $1\leq i\leq n$.
 Then, using the rule
 $R(f(x_1,\dots,x_n),f(y_1,\dots,y_n)) \to \True
 \Leftarrow R(x_1,y_1)\bowtie\True,\dots, R(x_n,y_n)\bowtie\True$,
 we get
 \begin{displaymath}
 \beta(T)\vdash_\mathrm{CRWL} 
   R(f(l_1',\dots,l_n'), f(r_1',\dots,r_n')) \to \True \,,
 \end{displaymath}
 verifying $f(l_1',\dots,l_n')\in [f(l_1,\dots,l_n)]$ and
 $f(r_1',\dots,r_n')\in [f(r_1,\dots,r_n)]$.
\item \textbf{Transitivity}. From
 \begin{displaymath}
 \bigfrac{[l]\to [t] \qquad [t]\to [r]}
         {[l]\to [r]}
 \end{displaymath}
 and the induction hypothesis, 
 $\beta(T)\vdash_\mathrm{CRWL} R(l', t')\to\True$ and
 $\beta(T)\vdash_\mathrm{CRWL} R(t'',r')\to\True$, with $l'\in [l]$,
 $t',t''\in [t]$, and $r'\in [r]$.
 Then, due to the equivalence proved above,
 $\beta(T)\vdash_\mathrm{CRWL} R(l, t) \to\True$
 and $\beta(T)\vdash_\mathrm{CRWL} R(t, r)\to\True$ and we get the 
 result using the translation of the \textbf{Transitivity} rule.
\item \textbf{Replacement}. We have, for some
 $[l(\overline{x})] \to [r(\overline{x})] \ifcond
 [a_1(\overline{x})]\to [b_1(\overline{x})]\land \dots \land
 [a_m(\overline{x})]\to [b_m(\overline{x})]$
 in $\Gamma$,
 \begin{displaymath}
 \bigfrac{\begin{array}{rcl}
           [w_1]\to [w_1']&\dots &[w_n]\to [w_n']\\
           {[}a_1(\overline{w}/\overline{x})] \to 
            [b_1(\overline{w}/\overline{x})] &\dots &%
            [a_m(\overline{w}/\overline{x})] \to  
            [b_m(\overline{w}/\overline{x})]
          \end{array}}
         {T\vdash_\mathrm{RL}[l(\overline{w}/\overline{x})] \to [r(\overline{w'}/\overline{x})]}.
 \end{displaymath}
 By induction hypothesis, there exist
 $a_i'\in [a_i(\overline{w}/\overline{x})],
  b_i'\in [b_i(\overline{w}/\overline{x})]$
 such that
 $\beta(T)\vdash_\mathrm{CRWL} R(a_i',b_i')\to\True$ for $i=1,\dots,m$.
 Again by the equivalence shown above,
 $\beta(T)\vdash_\mathrm{CRWL} R(a_i(\overline{w}/\overline{x}),b_i(\overline{w}/\overline{x}))\to\True$,
 for $i=1,\dots,m$.
 We can then use the linearised version of
 $R(l,r)\to\True \Leftarrow R(a_1,b_1)\bowtie \True,\dots,
  R(a_m,b_m)\bowtie \True$,
 substituting all variables which arose from the same one during
 the linearisation process with the same $w_i$
 (so that the conditions $x\bowtie x$, $x\bowtie y_j$ are
 trivially verified),
 to get $\beta(T)\vdash_\mathrm{CRWL}
 R(l(\overline{w}/\overline{x}),r(\overline{w}/\overline{x}))\to\True$.
 In a similar way,
 $\beta(T)\vdash_\mathrm{CRWL} R(w_i,w_i')\to\True$, $i=1,\dots,n$,
 is also obtained, and repeated aplication of the translation of
 the \textbf{Transitivity} and \textbf{Congruence} rules would show,
 first, that
 $\beta(T)\vdash_\mathrm{CRWL}
 R(r(\overline{w}/\overline{x}),r(\overline{w'}/\overline{x}))\to\True$,
 and then
 $\beta(T)\vdash_\mathrm{CRWL}
 R(l(\overline{w}/\overline{x}),r(\overline{w'}/\overline{x}))\to\True$,
 as desired. $\mathproofbox$
\end{itemize}
\end{proof*}

Our next goal will be to prove the converse of the last
proposition.
However, more care is needed here since, for example, an
equation of the form $x*0=0$ will allow us to derive
$R(\True*0,0)\to\True$.
Even more bizarre derivations are possible by repeated
application of transitivity, {e.g.}
$R(R(\True,\True),\True)\to\True$.
To prove that these rewrites, however,
do not allow us to derive anything in CRWL that was not already
derivable in the original RL-theory,
we concentrate first on some preliminary results.
The first one is proved by an easy induction for each fact in the
statement.
\begin{lemma}
Let $T=(\Sigma,E,\Gamma)$ be an RL-theory,
$\beta(T)=(\Sigma',\Gamma')$, and
$e, e_1,\dots, e_n\in \Bexpr{\Sigma'}{X}$
expressions in CRWL.
\begin{enumerate}
\item If\/ $\beta(T)\vdash_\mathrm{CRWL} \bot\to e$, then $e=\bot$.
\item For all $x\in\mathcal{X}$, if $\beta(T)\vdash_\mathrm{CRWL} x\to e$,
 then either $e=\bot$ or $e=x$.
\item If\/ $\beta(T)\vdash_\mathrm{CRWL} R(e_1,e_2)\to e$, then either
 $e=\bot$, or $e=\True$, or $e=R(e_1',e_2')$
 with $\beta(T)\vdash_\mathrm{CRWL} e_i\to e_i'$, $i=1,2$.
\item For every $f\in C_{\Sigma'}^n$, if
 $\beta(T)\vdash_\mathrm{CRWL} f(e_1,\dots,e_n)\to e$ then either
 $e=\bot$, or $e=f(e_1',\dots,e_n')$ with
 $\beta(T)\vdash_\mathrm{CRWL} e_i\to e_i'$ for some
 $e_i'\in\Bexpr{\Sigma'}{X}$, $i=1,\dots,n$. 
\end{enumerate}
\end{lemma}
In what follows this lemma will be used mostly without explicit
reference to it: for example, when deducing
$\beta(T)\vdash_\mathrm{CRWL} R(t,t')\to\True$ from
$\beta(T)\vdash_\mathrm{CRWL} R(t,t')\bowtie\True$.

\begin{lemma}
\label{lem:total}
Let $T$ be an RL-theory and $\beta(T)=(\Sigma',\Gamma')$.
\begin{enumerate}
\item \label{lem:totalfst}
 For all $e,e'\in\Bexpr{\Sigma'}{X}$,
 if $\beta(T)\vdash_\mathrm{CRWL} e\to e'$
 and $e'$ is total, then $e$ is total.
\item \label{lem:totalsnd}
 For all $t\in\Bterm{\Sigma'}{X}$, $e'\in\Bexpr{\Sigma}{X}$,
 if $\beta(T)\vdash_\mathrm{CRWL} t\to e'$
 and $e'$ is total, then $t=e'$.
\item \label{lem:totalthr}
 For all $t,t'\in\Bterm{\Sigma'}{X}$,
 if $\beta(T)\vdash_\mathrm{CRWL} t\bowtie t'$,
 then $t$ is total and $t'=t$.
\end{enumerate}
\end{lemma}
\begin{proof*}
\begin{enumerate}[3.]
\item By induction on the last rule of the derivation.
 Rules \textbf{Bottom} and \textbf{Join} are not possible, and 
 \textbf{Reflexivity} is immediate.
 For \textbf{Transitivity}, if the last step of the derivation is
 \begin{displaymath}
 \bigfrac{e\to e'' \qquad e''\to e'}{e\to e'}\,,
 \end{displaymath}
 then, by induction hypothesis, $e''$ is total, and again by induction
 hypothesis $e$ is total. 
 The situation is similar for \textbf{Monotonicity}.
 For \textbf{Reduction} we have to distinguish all these cases:
 \begin{itemize}
 \item If $R(x_1,x_2) \to \True \Leftarrow x_1 \bowtie x_2$ has been used,
  then $e=R(e_1,e_2)$ and $\beta(T)\vdash_\mathrm{CRWL} e_1\to t$,
  $\beta(T)\vdash_\mathrm{CRWL} e_2\to t$ for some $t$, total, have
  been previously obtained in the derivation.
  By induction hypothesis, both $e_1$ and $e_2$ are total and so is
  $e$.
 \item If a rule of the form
  $R(f(x_1,\dots,x_n),f(y_1,\dots,y_n)) \to \True
  \Leftarrow R(x_1,y_1)\bowtie \True,\dots, R(x_n,y_n)\bowtie\True$ or
  $R(x,y) \to \True \Leftarrow R(x,z)\bowtie\True, R(z,y)\bowtie\True$
  has been used, the result follows by induction hypothesis.
 \item If the last rule applied has been one of those corresponding to
  equations or rewrite rules then $e=R(l,r)$ and a condition of the form
  $x\bowtie x$ or $x\bowtie y_j$ for every variable appearing in it
  must have been satisfied.
  If $x$ has been instantiated with $t$, then those conditions imply
  that $\beta(T)\vdash_\mathrm{CRWL} t\to t'$ for some total $t'$,
  so by induction
  hypothesis $t$ is total, and so will be the expression $e$.
 \end{itemize}
\item By (1), $t\in\Term{\Sigma'}{X}$. By structural induction on $t$:
 \begin{itemize}
 \item $t=x$, then $e'=\bot$ (absurd) or $e'=x$ and the result holds;
 \item $t=f(t_1,\dots,t_n)$, then either $e'=\bot$ (absurd) or
  $e'=f(e_1',\dots,e_n')$ with $\beta(T)\vdash_\mathrm{CRWL} t_i\to e_i'$.
  In this last case, by induction hypothesis, $t_i=e_i'$ for
  $i=1,\dots,n$ and so $t=e'$.
 \end{itemize}
\item There exists $t''\in\Term{\Sigma'}{X}$ with
 $\beta(T)\vdash_\mathrm{CRWL} t\to t''$ and
 $\beta(T)\vdash_\mathrm{CRWL} t'\to t''$, and by (2), $t=t''=t'$.
 $\mathproofbox$
\end{enumerate}
\end{proof*}

We are now ready to prove our goal and we will do so in two steps.
The next proposition shows that if $R(l,r) \to \True$ can be 
proved in $\beta(T)$ then $[l] \to [r]$ can also be proved in $T$,
but extending the signature with the constant $\True$.
After that, we show that we can forget about this extra constant.
\begin{proposition}
\label{prop:RLinCRWLmainhalf}
Let $T=(\Sigma,E,\Gamma)$ be an RL-theory, $\beta(T)=(\Sigma',\Gamma')$,
and let $l,r \in\Bterm{\Sigma'}{X}$.
If $\beta(T)\vdash_\mathrm{CRWL} R(l,r)\to\True$ then
$l,r\in\Term{\Sigma'}{X}$ (recall that $\Term{\Sigma'}{X} =
T_{\Sigma\cup\{\mathit{true}\}}(\mathcal{X})$) and
$(\Sigma\cup\{\True\},E,$ $\Gamma)\vdash_\mathrm{RL} [l]\to [r]$.
\end{proposition}
\begin{proof*}
By Lemma~\ref{lem:total}.\ref{lem:totalfst}, $l,r\in\Term{\Sigma'}{X}$.
For the second part, we proceed
by induction on the proof of $\beta(T)\vdash_\mathrm{CRWL} R(l,r)\to\True$.
The last rule applied must have been \textbf{Transitivity} or 
\textbf{Reduction}.
\begin{itemize}
\item For \textbf{Transitivity} we have
 \begin{displaymath}
 \bigfrac{R(l,r)\to e \qquad e\to\True}
         {R(l,r)\to \True}.
 \end{displaymath}
 If $e=\True$ the result follows by induction hypothesis.
 Otherwise it must be $e=R(l',r')$, total by
 Lemma~\ref{lem:total}.\ref{lem:totalfst},
 with $\beta(T)\vdash_\mathrm{CRWL} l\to l'$ and
 $\beta(T)\vdash_\mathrm{CRWL} r\to r'$.
 Hence, by Lemma~\ref{lem:total}.\ref{lem:totalsnd}, $l=l'$,
 $r=r'$, and the result holds by induction hypothesis.
\item For \textbf{Reduction} there are five different cases,
 corresponding to each of the conditional rewrite rules simulating 
 the RL-calculus in $\beta(T)$.
 Recall that rules in CRWL are instantiated only with members 
 of \Bterm{\Sigma'}{X}.
 \begin{enumerate}
 \item If the last step of the derivation is
  \begin{displaymath}
  \bigfrac{l\bowtie r}
          {R(l,r)\to\True},
  \end{displaymath}
  then by Lemma~\ref{lem:total}.\ref{lem:totalthr} it is
  $l=r$ and therefore 
  $(\Sigma \cup \{\True\},E,\Gamma)\vdash_\mathrm{RL} [l]\to [r]$.
 \item If we have
  \begin{displaymath}
  \bigfrac{R(l,t)\bowtie\True \qquad R(t,r)\bowtie\True}
          {R(l,r)\to\True},
  \end{displaymath}
  then, by induction hypothesis,
  $(\Sigma \cup \{\True\},E,\Gamma)\vdash_\mathrm{RL} [l]\to [t]$ and
  $(\Sigma \cup \{\True\},E,\Gamma)\vdash_\mathrm{RL} [t]\to [r]$, so
  $(\Sigma \cup \{\True\},E,\Gamma)\vdash_\mathrm{RL} [l]\to [r]$
  by \textbf{Transitivity} of RL.
 \item For the translation of the \textbf{Congruence} rule the 
  result also follows immediately by the induction hypothesis.
 \item Assume that the result is obtained by using one of the
  (linearised) rules associated to an equation $t=t'\in E$.
  The conditions of the form $x\bowtie x$ and
  $x\bowtie y_j$ in the rule together with
  Lemma~\ref{lem:total}.\ref{lem:totalthr} imply that all the variables
  which arose from the same one must have been instantiated
  with the same element of \Term{\Sigma'}{X}.
  This way $E\vdash l=r$, so $[l] = [r]$ and
  $(\Sigma\cup \{\True\},E,\Gamma)\vdash_\mathrm{RL} [l]\to [r]$
  by \textbf{Reflexivity} of RL.
 \item If the last rule applied is one of those
  associated to an element of $\Gamma$
  then, as in the previous case,
  all variables have been instantiated properly
  and the result follows by the induction hypothesis and using
  \textbf{Replacement}. $\mathproofbox$
 \end{enumerate}
\end{itemize}
\end{proof*}

These results, combined with the completeness of RL, are enough to 
prove the converse of Proposition~\ref{prop:RLinCRWLhalf};
again, since the semantics of RL is not explained until 
Section~\ref{sec:TMRL}, we postpone the proof to the appendix.
\begin{proposition}
Given any RL-theory $T=(\Sigma,E,\Gamma)$, and
$l,r \in T_\Sigma(\mathcal{X})$:
\begin{displaymath}
\begin{array}{rcl}
T\vdash_\mathrm{RL} [l]\to[r] &\iff
 &(\exists l'\in[l],\exists r'\in [r]) \;\;
 \beta(T)\vdash_\mathrm{CRWL} R(l',r')\to\True  \\
&\iff &(\forall l'\in[l],\forall r'\in [r]) \;\;
 \beta(T)\vdash_\mathrm{CRWL} R(l',r')\to\True
\end{array}
\end{displaymath}
\end{proposition}

\section{Relations at the Semantic Level}

In this section we leave behind our study of the entailment 
relations and turn our attention to models and satisfaction.
Our interest lies in associating suitable institutions
to both CRWL and RL and, thereafter, to relate them via
maps of institutions with ``good'' properties.

\subsection{Institutions}

The notion of model is based on Goguen and Burstall's  work
on institutions \cite{GoguenBurstall92}.
An \emph{institution\/} is a 4-tuple
$\mathcal{I} = (\Sign,\Sen,\Mod,$ $\models)$ such that:
\begin{itemize}
\item
 \Sign\ is a category whose objects are called \emph{signatures}.
\item
 $\Sen: \Sign \to \Set$ is a functor associating to
 each signature $\Sigma$ a set of $\Sigma$-\emph{sentences}.
\item
 $\Mod: \Sign^\mathrm{op} \to \Cat$ is a functor that gives
 for each signature $\Sigma$ a category whose objects are called
 $\Sigma$-\emph{models}.
\item
 $\models$ is a function associating to each $\Sigma \in |\Sign|$ a
 binary relation
 $\models_{\Sigma} \; \subseteq |\Mod(\Sigma)| \times \Sen(\Sigma)$
 called $\Sigma$-\emph{satisfaction}, in such a way that the following
 property holds for any $H: \Sigma \to \Sigma'$,
 $M' \in |\Mod(\Sigma')|$, and all
 $\varphi \in \Sen(\Sigma)$:
 \begin{displaymath}
 M'\models_{\Sigma'} \Sen(H)(\varphi) \; \iff \;
 \Mod(H)(M') \models_{\Sigma} \varphi
 \end{displaymath}
\end{itemize}

Given a set $\Gamma$ of $\Sigma$-sentences, the category
$\Mod(\Sigma,\Gamma)$
is defined as the full subcategory of $\Mod(\Sigma)$ determined
by those models $M\in |\Mod(\Sigma)|$ that satisfy all the
sentences in $\Gamma$.
A relation between sets of sentences and sentences, also denoted as
$\models$, can be defined by
\begin{displaymath}
\Gamma\models_\Sigma \varphi \; \iff \; M\models_\Sigma \varphi
\ \textrm{for each $M\in|\Mod(\Sigma,\Gamma)|$} \,\textrm{.}
\end{displaymath}
We can then associate an entailment system to each institution
$\mathcal{I} = (\Sign, \Sen,$ $\Mod,\models)$ in a natural way
by means of the triple $\mathcal{I}^{+} = (\Sign,\Sen,\models)$,
where $\models$ now denotes the previously defined relation between
sets of sentences and sentences; $\mathcal{I}^{+}$ is easily seen to satisfy the
conditions to be an entailment system.

Given an institution $\mathcal{I}$, its category \Th\ of theories is
defined as the category of theories associated to the entailment
system $\mathcal{I}^{+}$.
If $H:(\Sigma,\Gamma)\to (\Sigma',\Gamma')$ is a theory morphism
and $M'\in \Mod(\Sigma',\Gamma')$, it is not difficult to check
that $\Mod(H)(M') \in \Mod(\Sigma,\Gamma)$.
The model functor \Mod\ can then be extended to a functor
$\Mod:\Th^\mathrm{op} \to \Cat$.

There are many different notions of morphisms between institutions 
in the literature; a good survey can be found in \citeN{GoguenRosu02}.
Although it will not play a crucial role in what follows, to give 
the reader a flavour of the idea we present here  the definition 
introduced in \citeN{Meseguer89}.
Given institutions $\mathcal{I} = (\Sign,\Sen,\Mod,\models)$
and $\mathcal{I}' = (\Sign',\Sen',\Mod',$ $\models')$, a
\emph{map of institutions} $(\Phi,\alpha,\beta):\mathcal{I} \to \mathcal{I}'$
consists of a natural transformation
$\alpha: \Sen \Rightarrow \Sen' \comp \Phi$, an $\alpha$-sensible
functor\footnote{Essentially, this means that $\Phi$ is 
determined by its restriction to empty theories and $\alpha$.}
$\Phi: \Th_0 \to \Th'_0$, and a natural
transformation $\beta: \Mod'\comp \Phi^\mathrm{op} \Rightarrow \Mod$ such
that for each $\Sigma \in |\Sign|$, $\varphi \in \Sen(\Sigma)$,
and $M' \in |\Mod'(\Phi(\Sigma,\emptyset))|$ the following property is
satisfied:
\begin{displaymath}
 M'\models'_{\Sigma'} \alpha_{\Sigma}(\varphi) \; \iff \;
\beta_{(\Sigma,\emptyset)}(M') \models_{\Sigma} \varphi
\end{displaymath}

\subsection{The models of RL} \label{sec:TMRL}

Before proceeding to $\mathcal{R}$-\emph{systems}, the models of
RL, we need the categorical notion of \emph{subequalizer}
\cite{Lambek70}, a notion generalizing
that of equalizer of two functors.\footnote{In \citeN{Miyoshi96},
subequalizers are shown to coincide with \emph{inserters}, a special
kind of weighted limit, in the 2-category \Cat. This allows the author
to generalize the models of RL, building them over arbitrary 2-categories
and even enriched categories.}

Given a family of pairs of functors $\{F_i, G_i: \mathcal{A}\to
\mathcal{B}_i \mid i \in I \}$, the (simultaneous) \emph{subequalizer}
of this family is a category $\textit{Subeq}((F_i,G_i)_{i\in I})$
together with a functor
\begin{displaymath}
J : \textit{Subeq}((F_i,G_i)_{i\in I}) \to \mathcal{A}
\end{displaymath}
and a family of natural transformations $\{\alpha_i: F_i \comp J \Rightarrow
G_i \comp J \mid i \in I\}$ satisfying the following universal property:
given a functor $H: \mathcal{C} \to \mathcal{A}$ and a family
of natural transformations $\{ \beta_i : F_i \comp H \Rightarrow G_i \comp H
\mid i\in I \}$, there exists a unique functor
$(H,\{\beta_i\}_{i\in I}) : \mathcal{C} \to \textit{Subeq}((F_i,G_i)_{i\in I})$
such that
\begin{displaymath}
J \comp (H,\{\beta_i \}_{i\in I}) = H \quad\textrm{and}\quad
\alpha_i \comp (H,\{\beta_i \}_{i\in I}) = \beta_i \quad (i\in I)\,\textrm{.}
\end{displaymath}

The construction of $\textit{Subeq}((F_i,G_i)_{i\in I})$ is quite
simple.
Its objects are pairs $(A,$ $\{ b_i\}_{i\in I})$ with $A$ an object
in $\mathcal{A}$ and $b_i: F_i(A) \to G_i(A)$  a morphism in
$\mathcal{B}_i$.
Morphisms $a : (A,\{b_i\}_{i\in I}) \to (A',\{b_i'\}_{i\in I})$
are morphisms $a:A \to A'$ in $\mathcal{A}$ such that for each
$i\in I$, $G_i(a)\comp b_i = b_i'\comp F_i(a)$.
The functor $J$ is just projection into the first component.
The natural transformations $\alpha_j$ are defined by
\begin{displaymath}
\alpha_j(A,\{b_i\}_{i\in I}) = b_j \quad (j\in I)\,\textrm{.}
\end{displaymath}

Then, given an RL-theory $\mathcal{R}= (\Sigma,E,L,\Gamma)$, an
$\mathcal{R}$-\emph{system} $\mathcal{S}$ is a category $\mathcal{S}$
together with:
\begin{itemize}
\item
 A $(\Sigma,E)$-algebra structure given by a family of functors
 \begin{displaymath}
 \{ f_\mathcal{S}: \mathcal{S}^n \to \mathcal{S} \mid
 f\in \Sigma \textrm{ of arity $n$} \}
 \end{displaymath}
 satisfying the equations $E$, i.e., for any
 $t(x_1,\dots,x_n) = t'(x_1,\dots,x_n)$ in $E$ we have an identity of
 functors $t_\mathcal{S} = t'_\mathcal{S}$, where the functor
 $t_\mathcal{S}$ is defined inductively from the functors
 $f_\mathcal{S}$ in the obvious way.
\item
 For each rewrite rule
 \begin{displaymath}
 r: [t(\overline{x})] \to [t'(\overline{x})] \ifcond
 [a_1(\overline{x})]\to [b_1(\overline{x})] \land\dots\land
 [a_m(\overline{x})]\to [b_m(\overline{x})]
 \end{displaymath}
 in $\Gamma$, a natural transformation
 \begin{displaymath}
 r_\mathcal{S}:
 t_\mathcal{S} \comp J_\mathcal{S} \Rightarrow
                t'_\mathcal{S} \comp J_\mathcal{S},
 \end{displaymath}
 where
 $J_\mathcal{S} : \textit{Subeq}((a_{j\mathcal{S}},
                                  b_{j\mathcal{S}})_{1\leq j\leq m})
                  \to \mathcal{S}^n$
 is the subequalizer functor.
\end{itemize}
An $\mathcal{R}$-\emph{homomorphism} $F:\mathcal{S} \to \mathcal{S}'$
between two $\mathcal{R}$-systems is then a functor
$F:\mathcal{S} \to \mathcal{S}'$ such that:
\begin{itemize}
\item It is a $\Sigma$-algebra homomorphism, i.e.,
 $F \comp f_\mathcal{S} = f_{\mathcal{S}'} \comp F^n$,
 for each $f$ in $\Sigma$ of arity $n$.
\item ``$F$ preserves $\Gamma$,'' i.e., for each rewrite rule
 $r: [t(\overline{x})] \to [t'(\overline{x})] \ifcond C$
 in $\Gamma$ we have the identity of natural transformations
 \begin{displaymath}
 F \comp r_\mathcal{S} = r_{\mathcal{S}'} \comp F^\bullet,
 \end{displaymath}
 where $F^\bullet : \textit{Subeq}(C_\mathcal{S}) \to
 \textit{Subeq}(C_{\mathcal{S}'})$ is the unique functor induced by
 the universal property of $\textit{Subeq}(C_{\mathcal{S}'})$ by the
 composition functor
 \begin{displaymath}
 \textit{Subeq}(C_\mathcal{S}) \stackrel{J_\mathcal{S}}{\longrightarrow}
  \mathcal{S}^n \stackrel{F^n}{\longrightarrow} \mathcal{S'}^n
 \end{displaymath}
 and the natural transformations $F \comp \alpha_j$, $1\leq j\leq m$,
 where the condition $C$ has $m$ rewrites $[a_j]\to [b_j]$, and
 $\alpha_j$ is the $j$th natural transformation associated
 to the subequalizer $\textit{Subeq}(C_\mathcal{S})$.
 Despite the somewhat complicated definition of $F^\bullet$, its
 behavior on objects is quite simple; it is given by the
 equation
 \begin{displaymath}
 F^\bullet(\overline{C}^n,\overline{c}^m) =
        (F^n(\overline{C}^n),F^m(\overline{c}^m))\,\textrm{.}
 \end{displaymath}
\end{itemize}
This defines a category $\mathcal{R}$-\textbf{Sys} in the obvious
way.

A sequent $[t(x_1,\dots,x_n)] \to [t'(x_1,\dots,x_n)]$ is satisfied
by
an $\mathcal{R}$-system $\mathcal{S}$ if there exists a natural
transformation
\begin{displaymath}
\alpha: t_\mathcal{S} \Rightarrow t'_\mathcal{S}
\end{displaymath}
between the functors
$t_\mathcal{S}, t'_\mathcal{S} :\mathcal{S}^n \to \mathcal{S}$.
We use the notation
\begin{displaymath}
\mathcal{S} \models [t(x_1,\dots,t_n)] \to [t'(x_1,\dots,x_n)]\,\textrm{.}
\end{displaymath}
With respect to this definition of satisfaction, the proof calculus
is sound and complete \cite{Meseguer92}.
Completeness is obtained by means of an initial model construction.

\subsection{The models of CRWL}

Before defining models we review some definitions.
A \emph{partially ordered set} (in short, poset)
with bottom $\bot$ is a set $S$
equipped with a partial order $\sqsubseteq$ and a least element $\bot$.
We say that an element $x\in S$ is \emph{totally defined}
if $x$ is maximal with respect to $\sqsubseteq$.
The set of all totally defined elements of $S$ will be denoted
$\textit{Def}(S)$.
$D\subseteq S$ is a \emph{directed} set if for all $x,y\in D$ there exists
$z\in D$ with $x\sqsubseteq z$, $y\sqsubseteq z$.
A subset $A\subseteq S$ is a \emph{cone} if $\bot\in A$ and, for all
$x\in A$ and $y\in S$, if $y\sqsubseteq x$ then $y\in A$.
An \emph{ideal} $I\subseteq S$ is a directed cone.
For $x\in S$, the principal ideal generated by $x$ is
$\langle x\rangle = \{y\in S \mid y\sqsubseteq x\}$.
We write $\mathcal{C}(S)$ for the set of cones of $S$.

Given a signature $\Sigma$, a \emph{CRWL-algebra} over
$\Sigma$ is a triple
\begin{displaymath}
\mathcal{A} = (D^\mathcal{A},\{c^\mathcal{A}\}_{c\in C_\Sigma},\{f^\mathcal{A}\}_{f\in F_\Sigma}),
\end{displaymath}
where $D^\mathcal{A}$ is a poset with bottom, and $c^\mathcal{A}$
and $f^\mathcal{A}$ are monotone mappings from $(D^\mathcal{A})^n$
to $\mathcal{C}(D^\mathcal{A})$, with $n$ the corresponding arity.
In addition, for $c\in C_\Sigma^n$ and
for all $u_1,\dots,u_n\in D^\mathcal{A}$, there
exists a $v\in D^\mathcal{A}$ such that
$c^\mathcal{A}(u_1,\dots,u_n) = \langle v\rangle$.
Moreover, $v\in \textit{Def}(D^\mathcal{A})$ in case that all
$u_i\in \textit{Def}(D^\mathcal{A})$.

Note that any $h: S\to \mathcal{C}(S')$ can be extended to a function
$\hat{h}: \mathcal{C}(S) \to \mathcal{C}(S')$ defined by
$\hat{h}(x) = \bigcup_{x\in S} h(x)$.
By abuse of notation, we will write $\hat{h}$ also as $h$.

A \emph{valuation} over $\mathcal{A}$ is any mapping
$\eta:\mathcal{X}\to D^\mathcal{A}$, and we say that $\eta$ is
\emph{totally defined} if $\eta(x)\in\textit{Def}(D^\mathcal{A})$
for all $x \in\mathcal{X}$.
The \emph{evaluation} of an expression $e\in\Bexpr{\Sigma}{X}$ in
$\mathcal{A}$ under $\eta$ yields
$\lsem e\rsem^\mathcal{A}\eta\in\mathcal{C}(D^\mathcal{A})$,
which is defined recursively as follows:
\begin{itemize}
\item $\lsem\bot\rsem^\mathcal{A}\eta = \langle\bot_\mathcal{A}\rangle$.
\item $\lsem x\rsem^\mathcal{A}\eta = \langle \eta(x)\rangle$,
 for $x\in\mathcal{X}$.
\item $\lsem h(e_1,\dots,e_n)\rsem^\mathcal{A}\eta =
       h^\mathcal{A}(\lsem e_1\rsem^\mathcal{A}\eta,\dots,
                     \lsem e_n\rsem^\mathcal{A}\eta)$,
 for all $h\in C_\Sigma^n\cup F_\Sigma^n$.
\end{itemize}

We are now prepared to define models.
Let $\mathcal{A}$ be CRWL-algebra $\mathcal{A}$:
\begin{itemize}
\item $\mathcal{A}$ satisfies a reduction statement $a\to b$ under
 a valuation $\eta$, $(\mathcal{A},\eta)\models a\to b$, if
 $\lsem a\rsem^\mathcal{A}\eta \supseteq\lsem b\rsem^\mathcal{A}\eta$.
\item $\mathcal{A}$ satisfies a joinability statement $a\bowtie b$
 under $\eta$, $(\mathcal{A},\eta)\models a\bowtie b$, if
 $\lsem a\rsem^\mathcal{A}\eta \cap\lsem b\rsem^\mathcal{A}\eta
  \cap \textit{Def}(D^\mathcal{A}) \neq\emptyset$.
\item $\mathcal{A}$ satisfies a rule $l\to r\Leftarrow C$ if every
 valuation $\eta$ such that $(\mathcal{A},\eta)\models C$ verifies
 $(\mathcal{A},\eta)\models l\to r$.
\item $\mathcal{A}$ is a \emph{model} of $\Gamma$, $\mathcal{A}\models
 \Gamma$ if $\mathcal{A}$ satisfies all the rules in $\Gamma$.
\end{itemize}
As mentioned in Section~\ref{sec:AESCRWL}, the CRWL-calculus 
is \emph{partially} sound and complete \cite{GonzalezEtAl99}
with respect to this notion of satisfaction:
\begin{itemize}
\item If $\varphi$ is a reduction or a joinability statement,
 $\Gamma\vdash_\mathrm{CRWL}\varphi$ implies that
 $(\mathcal{A},\eta) \models \varphi$, for
 every $\mathcal{A}\models\Gamma$ and every \emph{totally defined}
 valuation $\eta$.
\item If $\varphi$ is a joinability statement or a reduction statement in
 which the righthand expression is a partial term,
 the previous implication becomes an equivalence.
\end{itemize}

Finally, we can also define homomorphisms between CRWL-algebras.
Let $\mathcal{A}$, $\mathcal{B}$ be two CRWL-algebras over a
signature $\Sigma$.
A \emph{CRWL-homomorphism} $H:\mathcal{A}\to \mathcal{B}$ is a
monotone function $H:D^\mathcal{A}\to \mathcal{C}(D^\mathcal{B})$
which satisfies the following conditions:
\begin{enumerate}
\item $H$ is element-valued:
 for all $u\in D^\mathcal{A}$ there exists $v\in D^\mathcal{B}$
 such that $H(u)= \langle v\rangle$.
\item $H$ is strict:
 $H(\bot_\mathcal{A}) = \langle\bot_\mathcal{B}\rangle$.
\item $H$ preserves constructors: for all $c\in C_\Sigma^n$,
 $u_i\in D^\mathcal{A}$, is $H(c^\mathcal{A}(u_1,\dots,u_n)) =
  c^\mathcal{B}(H(u_1),\dots,H(u_n))$.
\item $H$ loosely preserves defined functions: that is,
 for all $f\in F_\Sigma^n$,
 $u_i\in D^\mathcal{A}$, $H(f^\mathcal{A}(u_1,\dots,u_n))$ $\subseteq
  f^\mathcal{B}(H(u_1),\dots,H(u_n))$.
\end{enumerate}
CRWL-algebras as objects with CRWL-homomorphisms as arrows form
a category.

\subsection{An institution for CRWL}
\label{sec:ICRWL}

An institution for CRWL was first defined in \citeN{Molina-thesis}.
This institution, however, was defined with the goal of providing a
basis for the semantics of modules in CRWL, and restricts its
attention to a class of particular term algebras.
Since our goal is more general, we do not place such a
limitation and define
$\mathcal{I}_\mathrm{CRWL}=(\Sign,\Sen,\Mod,\models)$
as follows:
\begin{itemize}
\item \Sign: the category of signatures with constructors and
 signature morphisms.
\item $\Sen:\Sign\to\Set$ the functor assigning to each signature
 $\Sigma$ the set of all conditional rewrite rules over it, and
 to each signature morphism $\sigma$ its homomorphic extension 
 to rewrite rules, with $\sigma(\bot) = \bot$.
\item $\Mod:\Sign^\mathrm{op}\to\Cat$ the functor assigning
 to each signature the category of CRWL-algebras and homomorphisms
 over it, and to
 each $\sigma:\Sigma\to\Sigma'$ the forgetful functor
 mapping $\mathcal{A}'\in |\Mod(\Sigma')|$ to the CRWL-algebra
 $\mathcal{A}'_\sigma$
 with the same underlying poset and such that
 $h^{\mathcal{A}'_\sigma} = \sigma(h)^{\mathcal{A}'}$ for all
 $h\in\Sigma$, and which is the identity over homomorphisms.
\item $\models$ the satisfaction relation in CRWL.
\end{itemize}

\begin{proposition}
\label{prop:ICRWL}
$\mathcal{I}_\mathrm{CRWL}$ is an institution.
\end{proposition}
\begin{proof}
It is not difficult to check that \Sign\ is a category, and
that \Sen\ and \Mod\ are indeed functors.
As for the satisfaction condition, let $\sigma:\Sigma \to \Sigma'$
be a signature morphism, $\mathcal{A}'\in |\Mod(\Sigma')|$,
and $\varphi\in \Sen(\Sigma)$; we have to prove that
\[
\mathcal{A}'\models \sigma(\varphi) \iff
\mathcal{A}'_\sigma\models \varphi \,\textrm{.}
\] 
It is easy to show, by structural induction on $e$, that
\[
\lsem e \rsem^{\mathcal{A}'_\sigma}\eta =
\lsem \sigma(e) \rsem^{\mathcal{A}'}\eta
\]
for every $e\in \Bexpr{\Sigma}{\mathcal{X}}$ and valuation
$\eta$ over $\mathcal{A}'$. 
Let $\varphi = e \to e'$ be a reduction statement.
Then, for any valuation $\eta$,
\[
\begin{array}{ccccc}
(\mathcal{A}',\eta)\models \sigma(\varphi) &\iff& 
     \lsem \sigma(e')\rsem^{\mathcal{A}'}\eta \subseteq
     \lsem \sigma(e)\rsem^{\mathcal{A}'}\eta \\
&\iff & \lsem e'\rsem^{\mathcal{A}'_\sigma}\eta \subseteq
        \lsem e\rsem^{\mathcal{A}'_\sigma} \eta
      &\iff & (\mathcal{A}'_\sigma,\eta)\models \varphi            
\end{array}
\]
and analogously for $\varphi$ a joinability statement.
Now, if $l\to r \Leftarrow C$ is a conditional rewrite 
rule, it follows that $\mathcal{A}'_\sigma\models C \iff
\mathcal{A}'\models \sigma(C)$ and 
$\mathcal{A}'_\sigma\models l\to r \iff 
\mathcal{A}'\models \sigma(l\to r)$,
and thus the satisfaction condition is indeed verified.
\end{proof}

It can be proved that the category $\Mod(T)$ has products for every
CRWL-theory $T$; it is not complete, however, as 
in Section~\ref{sec:SE}
it is shown that, in general, $\Mod(T)$ does not have equalizers.
${\mathcal I}_\mathrm{CRWL}$ is also a semiexact 
institution \cite{Palomino-mthesis}.

\subsection{An institution for RL}
\label{sec:IRL}

The task of assigning an institution to RL is harder than expected.
The first and most natural idea is to define the category of
signatures \Sign\ as the category of equational theories and theory
morphisms, and the functor \Sen\ to map any such theory to the set
of conditional rewrite rules over it.
Since there are also notions of model and satisfaction in RL,
the desired institution seems to be at hand.
However, when one tries to put together the various components
of the institution, problems start to arise.

In the first place, the notion of satisfaction in RL is defined only
for unconditional rewrite rules, so our first task must be to extend
its definition so as to encompass the conditional ones.
Taking the definition of $\mathcal{R}$-systems as a guide, we say
that a conditional rewrite rule
\begin{displaymath}
[t(\overline{x})]\to [t'(\overline{x})] \ifcond
 [a_1(\overline{x})]\to [b_1(\overline{x})] \land\dots \land
 [a_m(\overline{x})]\to [b_m(\overline{x})]
\end{displaymath}
is satisfied by an $\mathcal{R}$-system $\mathcal{S}$ if there
exists a natural transformation
\begin{displaymath}
\alpha : t_\mathcal{S} \comp J_\mathcal{S} \Rightarrow
         t'_\mathcal{S} \comp J_\mathcal{S},
\end{displaymath}
where
$J_\mathcal{S} : \textit{Subeq}((a_{j\mathcal{S}},
                                 b_{j\mathcal{S}})_{1\leq j\leq m})
                 \to \mathcal{S}^n$.
(Alternatively, one could also think of defining satisfaction by
\begin{displaymath}
\mathcal{S} \models [t] \to [t'] \ifcond
[a_1] \to [b_1] \land \dots \land [a_m]\to [b_m]
\end{displaymath}
if
\begin{displaymath}
\mathcal{S} \models [a_i] \to [b_i]
\;\;\; i=1,\dots,m \quad \Longrightarrow \quad
\mathcal{S} \models [t] \to [t']\,\textrm{.}
\end{displaymath}
This option looks natural, but it is too loose in the sense
that it requires the consequent to hold only if the condition
is true for \emph{all} possible instances.
Note that, in our definition, the subequalizer is playing the same
role valuations have in the definition of satisfaction in CRWL.)

We can now prove the following proposition, that justifies the
soundness and completeness of the extended RL-calculus
presented in Section~\ref{sec:ESRLCRWL}.
\begin{proposition}
Let $\mathcal{R}$ be an RL-theory and
$[t(\overline{x})]\to [t'(\overline{x})] \ifcond
 [a_1(\overline{x})] \to [b_1(\overline{x})]\land \dots \land
 [a_m(\overline{x})] \to [b_m(\overline{x})]$ a conditional rewrite rule;
then, the following statements are equivalent:
\begin{enumerate}
\item $\mathcal{R} \models [t(\overline{x})]\to [t'(\overline{x})]
 \ifcond
 [a_1(\overline{x})] \to [b_1(\overline{x})]\land \dots \land
 [a_m(\overline{x})] \to [b_m(\overline{x})]$;
\item $\mathcal{R}(\overline{x}) \cup
 \{ [a_1(\overline{x})] \to [b_1(\overline{x})], \dots,
    [a_m(\overline{x})] \to [b_m(\overline{x})] \}
  \models [t(\overline{x})] \to [t'(\overline{x})]$;
\item $\mathcal{R}(\overline{x}) \cup
 \{ [a_1(\overline{x})] \to [b_1(\overline{x})], \dots,
    [a_m(\overline{x})] \to [b_m(\overline{x})] \}
  \vdash [t(\overline{x})]\to [t'(\overline{x})]$.
\end{enumerate}
\end{proposition}
\begin{proof*}
Statements (2) and (3) are equivalent by the soundness and completeness
of the RL-calculus \cite{Meseguer92}.
We will now prove that (1) implies (2) and then, that (3) implies (1).

To see that (1) implies (2), let $\mathcal{S}$ be an
$\mathcal{R}(\overline{x}) \cup
 \{ [a_1(\overline{x})] \to [b_1(\overline{x})], \dots,
    [a_m(\overline{x})] \to [b_m(\overline{x})] \}$-system.
There exist, therefore, natural transformations
\begin{displaymath}
h_j: a_j(\overline{x})_\mathcal{S} \to
     b_j(\overline{x})_\mathcal{S} 
\end{displaymath}
for $j=1,\dots,m$.
Since in this context, that is, over $\Sigma(\overline{x})$, 
both $t(\overline{x})$ and $t'(\overline{x})$
(as well as all the $a_j(\overline{x})$ and $b_j(\overline{x})$)
are ground terms, we only need to find a morphism 
$t(\overline{x})_\mathcal{S} \to t'(\overline{x})_\mathcal{S}$
in $\mathcal{S}$ to prove that 
$\mathcal{S} \models [t(\overline{x})] \to [t'(\overline{x})]$,
and it turns out that each $h_j$ is just a single morphism.
Let us write $\mathcal{S}|_\Sigma$ for the restriction of
$\mathcal{S}$ to the signature $\Sigma$ (that is, 
$\mathcal{S}|_\Sigma$ is like $\mathcal{S}$ but we forget the
interpretations for $\overline{x}$). 
Clearly, $\mathcal{S}|_\Sigma$ is an $\mathcal{R}$-system
and therefore, by hypothesis, there exists a natural transformation
\begin{displaymath}
\alpha : t_{\mathcal{S}|_\Sigma} \comp J_{\mathcal{S}|_\Sigma} 
         \Rightarrow
         t'_{\mathcal{S}|_\Sigma} \comp J_{\mathcal{S}|_\Sigma},
\end{displaymath}
where
$J_{\mathcal{S}|_\Sigma} : 
  \textit{Subeq}((a_{j\mathcal{S}|_\Sigma},
                  b_{j\mathcal{S}|_\Sigma})_{1\leq j\leq m})
  \to {\mathcal{S}|_\Sigma}^n$.
Because of the $h_j$, $1\leq j\leq m$, and noting that
$a_j(\overline{x})_\mathcal{S} = 
 a_{j\mathcal{S}|_\Sigma}(\overline{x}_\mathcal{S})$
(and analogously for $b_j$), the interpretation $\overline{x}_\mathcal{S}$
of the variables
$\overline{x}$ in $\mathcal{S}$ belongs to the subequalizer:
$(\overline{x}_\mathcal{S},\overline{h}) \in
            \textit{Subeq}((a_{j\mathcal{S}|_\Sigma},
                            b_{j\mathcal{S}|_\Sigma})_{1\leq j\leq m})$.
But then $\alpha(\overline{x}_\mathcal{S},\overline{h})$ is a
morphism 
$t_{\mathcal{S}|_\Sigma}(\overline{x}_\mathcal{S}) \to 
 t'_{\mathcal{S}|_\Sigma}(\overline{x}_\mathcal{S})$
in $\mathcal{S}|_\Sigma$, and therefore a morphism 
$t(\overline{x})_\mathcal{S} \to t'(\overline{x})_\mathcal{S}$
in $\mathcal{S}$, as required.

To show that (3) implies (1),
given an $\mathcal{R}$-system $\mathcal{S}$ we will prove
by induction on the derivation that
\begin{displaymath}
\mathcal{S}\models [t(\overline{x})]\to [t'(\overline{x})] 
 \ifcond
 [a_1(\overline{x})] \to [b_1(\overline{x})]\land \dots \land
 [a_m(\overline{x})] \to [b_m(\overline{x})]\,\textrm{.}
\end{displaymath}
According to the last rule of deduction employed:
\begin{itemize}
\item \textbf{Reflexivity}. It must be $[t] = [t']$ and the result 
 is immediate.
\item \textbf{Congruence}. If the last step in the derivation is
 \begin{displaymath}
 \bigfrac{[t_1] \to [t'_1] \;\;\; \dots \;\;\;
          [t_p]\to [t'_p]}
         {[f(t_1,\dots,t_p)]\to [f(t'_1,\dots,t'_p)]},
 \end{displaymath}
 we have, by the induction hypothesis,
 \begin{displaymath}
 \mathcal{S}\models [t_i]\to [t'_i] \ifcond
  [a_1] \to [b_1]\land \dots \land [a_m] \to [b_m] \qquad 1\leq i\leq p,
 \end{displaymath}
 and there exist natural transformations
 $\alpha_i : t_{i\mathcal{S}} \comp J_\mathcal{S} \Rightarrow
             t'_{i\mathcal{S}} \comp J_\mathcal{S}$, $1\leq i\leq p$,
 where $J_\mathcal{S} : \textit{Subeq}((a_{j\mathcal{S}},
                                        b_{j\mathcal{S}})_{1\leq j\leq m})
                        \to \mathcal{S}^n$.                        
 Let $(\overline{s},\overline{m}) \in
      \textit{Subeq}((a_{j\mathcal{S}},b_{j\mathcal{S}})_{1\leq j\leq m})$;
 if we define
 \begin{displaymath}
 \alpha(\overline{s},\overline{m}) =
          f_\mathcal{S}(\alpha_1(\overline{s},\overline{m}), \dots,
                        \alpha_p(\overline{s},\overline{m})),
 \end{displaymath}
 we obtain a natural transformation
 $\alpha : f(t_1,\dots,t_p) \comp J_\mathcal{S} \Rightarrow
           f(t'_1,\dots,t'_n) \comp J_\mathcal{S}$
 and the result is proved.
 Some warning words are in order here.
 In the functor $J_\mathcal{S} : \textit{Subeq}((a_{j\mathcal{S}},
                                     b_{j\mathcal{S}})_{1\leq j\leq m})
                     \to \mathcal{S}^n$,
 the $n$ appearing as superscript depends on the actual number
 of variables in the sentence
 $[t_i]\to [t'_i] \ifcond
  [a_1] \to [b_1]\land \dots \land [a_m] \to [b_m]$
 and, although the $a_j$ and $b_j$ are fixed, this is not the
 case for $t_i$ and $t'_i$ and thus the $n$ may vary with each $i$.
 This would imply that the category
 $\textit{Subeq}((a_{j\mathcal{S}}, b_{j\mathcal{S}})_{1\leq j\leq m})$
 could vary as well, since its objects are pairs whose first component is 
 an  object of $\mathcal{S}^n$, and then the definition of $\alpha$ 
 given above would no longer be valid.
 However, this is only a technical nuisance because the extra variables
 that $t_i$ and $t'_i$ may add are simply ignored by the functors
 $a_{j\mathcal{S}}$ and $b_{j\mathcal{S}}$, and everything could be
 made to fit properly by using projection functors that would
 preserve the natural transformations.
 This same remark applies to the remaining cases, too.
\item \textbf{Transitivity}. If we have
 \begin{displaymath}
 \bigfrac{[t] \to [t'] \qquad [t'] \to [t'']}
         {[t] \to [t'']},
 \end{displaymath}
 by induction hypothesis there exist natural transformations
 \begin{displaymath}
 \alpha_1 : t_\mathcal{S}\comp J_\mathcal{S} \Rightarrow
            t'_\mathcal{S} \comp J_\mathcal{S}
 \quad\textrm{and}\quad
 \alpha_2 : t'_\mathcal{S}\comp J_\mathcal{S} \Rightarrow
            t''_\mathcal{S} \comp J_\mathcal{S},
 \end{displaymath}
 where $J_\mathcal{S} : \textit{Subeq}((a_{j\mathcal{S}},
                                        b_{j\mathcal{S}})_{1\leq j\leq m})
                        \to \mathcal{S}^n$;
 the composition $\alpha_2 \comp \alpha_1$ gives the result.
\item \textbf{Replacement}. We distinguish two cases:
 \begin{enumerate}
 \item The rule employed is one of the
  $[a_j(\overline{x})] \to [b_j(\overline{x})]$.
  Since the terms are ground we must have
  \begin{displaymath}
  \bigfrac{}{[a_j(\overline{x})] \to [b_j(\overline{x})]}.
  \end{displaymath}
  But in this case,
  $\mathcal{S}\models [a_i]\to [b_i] \ifcond
   [a_1] \to [b_1] \land \dots \land
   [a_m] \to [b_m]$ follows because the construction of the
  subequalizer produces a natural transformation
  $\alpha_j : a_{j\mathcal{S}} \comp J_\mathcal{S} \Rightarrow
              b_{j\mathcal{S}} \comp J_\mathcal{S}$.
 \item For some rule
  $[l(\overline{y})] \to [r(\overline{y})] \ifcond
   [u_1(\overline{y})] \to [v_1(\overline{y})] \land \dots \land
   [u_q(\overline{y})] \to [v_q(\overline{y})]$
  in $\mathcal{R}$, we have
  \begin{displaymath}
  \bigfrac{{[w_1] \to [w'_1] \;\;\; \dots \;\;\;
            [w_p] \to [w'_p]
            \atop
            [u_1(\overline{w}/\overline{y})] \to [v_1(\overline{w}/\overline{y})]
            \;\;\; \dots \;\;\;
            [u_q(\overline{w}/\overline{y})] \to [v_q(\overline{w}/\overline{y})]}}
          {[l(\overline{w}/\overline{y})] \to [r(\overline{w'}/\overline{y})]}.
  \end{displaymath}
  By the induction hypothesis there exist natural transformations
  \begin{displaymath}
  \alpha_i : w_{i\mathcal{S}} \comp J_\mathcal{S} \Rightarrow
             w'_{i\mathcal{S}} \comp J_\mathcal{S} \qquad 1\leq i\leq p,
  \end{displaymath}
  and
  \begin{displaymath}
  \beta_i : u_i(\overline{w})_\mathcal{S} \comp J_\mathcal{S} 
            \Rightarrow
            v_i(\overline{w})_\mathcal{S} \comp J_\mathcal{S} 
            \qquad 1\leq i\leq q,
  \end{displaymath}
  where $J_\mathcal{S} : 
               \textit{Subeq}((a_{j\mathcal{S}},
                               b_{j\mathcal{S}})_{1\leq j\leq m})
               \to \mathcal{S}^n$.
  Since $\mathcal{S}$ is an $\mathcal{R}$-system,
  there also exists a natural transformation
  \begin{displaymath}
  \gamma : l_\mathcal{S} \comp J'_\mathcal{S} \Rightarrow
           r_\mathcal{S} \comp J'_\mathcal{S}
  \end{displaymath}
  where $J'_\mathcal{S} : 
               \textit{Subeq}((u_{j\mathcal{S}},
                              v_{j\mathcal{S}})_{1\leq j\leq q})
               \to \mathcal{S}^p$.
  We now need to find a natural transformation
  $\alpha : l(\overline{w})_\mathcal{S} \comp J_\mathcal{S} \Rightarrow
            r(\overline{w'})_\mathcal{S} \comp J_\mathcal{S}$.
  For that, let $(\overline{s},\overline{m})$ be an object in
  $\textit{Subeq}((a_{j\mathcal{S}},b_{j\mathcal{S}})_{1\leq j\leq m})$;
  due to the morphisms $\beta_i(\overline{s},\overline{m})$ it turns out
  that $(\overline{w_\mathcal{S}(\overline{s})},
         \overline{\beta(\overline{s},\overline{m})})$ belongs
  to $\textit{Subeq}((u_{j\mathcal{S}}, v_{j\mathcal{S}})_{1\leq j\leq q})$
  and we can define
  \begin{displaymath}
  \alpha(\overline{s},\overline{m}) =
      r_\mathcal{S}\big(\overline{\alpha(\overline{s},\overline{m})}\big)
      \comp \gamma(\overline{w_\mathcal{S}(\overline{s})},
                   \overline{\beta(\overline{s},\overline{m})}),
  \end{displaymath}
  which finishes the proof. $\mathproofbox$
 \end{enumerate}
\end{itemize}
\end{proof*}

A more serious problem, as far as the definition of an institution 
for RL is concerned, is posed by the functor
$\Mod:\Sign^\mathrm{op} \to \Cat$ mapping signatures to models.
The difficulty resides in the fact that, in RL, models are 
assigned directly to RL-theories instead of signatures, as it
is customary in other logics.
One obvious solution would be to consider a signature $(\Sigma,E)$ as
a theory $\mathcal{R}=(\Sigma,E,\emptyset,\emptyset)$ with empty
set of axioms (and labels), and to map $(\Sigma,E)$ to the category
$\mathcal{R}$-\textbf{Sys} of models of $\mathcal{R}$.
But this approach presents an important drawback. 
Up to this point in the paper, we have omitted any explicit mention
of the set of labels of an RL-theory.
Although this was a safe convention when talking about deduction,
it is no longer the case when our interest shifts to models.
Thanks to the set of labels $L$ in an RL-theory
$\mathcal{R}=(\Sigma,E,L,\Gamma)$, the elements of $\Gamma$
become special, \emph{labeled} rewrite rules.
These rules force $\mathcal{R}$-systems to have a certain internal
structure: not only must $\mathcal{R}$-systems satisfy them,
but must also associate to them a \emph{distinguished} interpretation
(natural transformation) that must be preserved by 
homomorphisms. 
(In particular, the same rule could appear \emph{twice} in an 
RL-theory $\mathcal{R}$ under two different labels.
$\mathcal{R}$-systems are then forced to provide two, possibly
different, interpretations for the same rule, each of them to be
preserved by the homomorphisms.)
When considering a signature as a theory with empty sets of axioms,
we are not taking into account labeled rewrite rules.
This way, homomorphisms are not subjected to preserve any rewrite
rule and the categories $\Mod(\Gamma)$ of models of $\Gamma$ and 
$\mathcal{R}$-\textbf{Sys} of $\mathcal{R}$-systems, that we
expected to be the same, turn out to be different.

In \citeN{Palomino-mthesis}, some others attempts at defining
an institution with $\Sign$ as the category of equational theories
are explored but, since they cannot reflect the distinction between
labeled rules belonging to RL-theories
and unlabeled rules, all of them are bound to failure.
For this reason we are led to an institution in which the category
\Sign\ subsumes all the information of an RL-theory.
More precisely, we define
$\mathcal{I}_\mathrm{RL} = (\Sign,\Sen,\Mod,\models)$ where:
\begin{itemize}
\item \Sign\ is the \emph{discrete} category of RL-theories.
\item $\Sen:\Sign \to \Set$ maps each RL-theory to the set
 of conditional rewrite rules that can be built over its signature.
\item $\Mod: \Sign^\mathrm{op} \to \Cat$ maps an RL-theory
 $\mathcal{R}$ to the category $\mathcal{R}$-\textbf{Sys}.
\item $\models$ the satisfaction relation conveniently
 extended to conditional rewrite rules as discussed above.
\end{itemize}
Since \Sign\ is discrete, this trivially defines an institution.
Admittedly, this restriction seems to be
not justified.
In fact, two types of morphisms of RL-theories are proposed
in \citeN{Meseguer90}.
Basically, they are morphisms of equational theories ``preserving''
the rules in the RL-theories.
For our purposes, however, the present definition is general enough 
as it stands and its extension would not modify the use we will
make of it in the next section.

There exist other institutions associated to (variants of) RL
in the literature, e.g., \cite{Cengarle98,DiaconescuFutatsugi02};
in these papers, the objects in the category of signatures are the
sets of function symbols, without any rules.\label{preorder:pag}
As a consequence of this simplicity and the reasons we have mentioned above, 
the general categorical models of RL must
be somehow restricted and the choice in these two works is 
to require them to be preorders instead of arbitrary categories.

\subsection{Searching for embeddings}
\label{sec:SE}

Now that we have institutions associated to both RL and CRWL
capturing formally their semantics, we would like to relate 
them by means of maps of institutions having ``nice properties.''
In particular, due to the generality of RL and its 
$\mathcal{R}$-systems, a natural question to ask is whether 
$\mathcal{I}_\mathrm{CRWL}$ can be considered as a subinstitution
of $\mathcal{I}_\mathrm{RL}$.

The formal definition of subinstitution appeared originally
in \citeN{Meseguer89} and has been further generalized in subsequent
works.
One of those extensions was introduced in \citeN{Meseguer98},
where it is called an \emph{embedding}.
Embeddings are very general: the only requirement they impose on 
a map of institutions
$(\Phi,\alpha,\beta):\mathcal{I}\to\mathcal{I}'$
is that for each $T\in |\Th_\mathcal{I}|$,
the functor $\beta_T:\Mod'(\Phi(T))\to \Mod(T)$ has to be
an equivalence of categories.

We will show, however, that there is no
embedding from $\mathcal{I}_\mathrm{CRWL}$ into
$\mathcal{I}_\mathrm{RL}$.
For that, it will be enough to find a categorical property
which is preserved by an equivalence of categories and a
theory $T\in|\Th_\mathrm{CRWL}|$ such that 
$\Mod_\mathrm{RL}(\Phi(T))$ satisfies it whereas
$\Mod_\mathrm{CRWL}(T)$ does not.

Let $\Sigma$ be a signature with constructors such that
$C_\Sigma=\emptyset$ and $F_\Sigma$ consists of just two
constants $f_1$ and $f_2$,
$\Gamma = \{ f_2 \to x \Leftarrow f_1\bowtie f_1\}$,
and consider the CRWL-theory $T=(\Sigma,\Gamma)$.
We define two CRWL-algebras over $\Sigma$:
$\mathcal{A}$ given by the set $D^{\mathcal A}= \{\bot, a_1, a_2\}$
with partial order $\bot \sqsubseteq a_1 \sqsubseteq a_2$, and
the cones $f_1^{\mathcal A} = \langle a_1\rangle$ and
$f_2^\mathcal{A}= \langle\bot \rangle$;
and ${\mathcal B}$ with $D^{\mathcal B} = \{ \bot, b_1\}$, and
the cones $f_1^{\mathcal B} = f_2^{\mathcal B} = \langle \bot\rangle$.
$\mathcal{A},\mathcal{B}\in |\Mod_\mathrm{CRWL}(T)|$ trivially,
because they
do not satisfy the condition $f_1\bowtie f_1$.

Let us now define two CRWL-homomorphisms $F, G:{\mathcal A}\to {\mathcal B}$,
given by:
\begin{displaymath}
\begin{array}{ccc}
F(x) = \langle\bot\rangle &
\textrm{ and } &
G(x) = \left\{ \begin{array}{ll}
         \langle\bot\rangle  &\textrm{if $x=\bot, a_1$} \\
         \langle b_1\rangle  &\textrm{if $x=a_2$.}
       \end{array} \right.
\end{array}
\end{displaymath}
Clearly, $F$ and $G$ preserve both $f_1$ and $f_2$, so that they are
actually homomorphisms;
we will prove that there is no equalizer of $F$ and $G$.
For let us assume that $E:{\mathcal E}\to {\mathcal A}$ is such an 
equalizer and let $H:{\mathcal A}\to {\mathcal A}$ be the homomorphism 
given by
\begin{displaymath}
H(x) = \left\{ \begin{array}{ll}
        \langle\bot\rangle & \textrm{if $x=\bot$} \\
        \langle a_1\rangle & \textrm{if $x=a_1, a_2$,}
       \end{array} \right.
\end{displaymath}
satisfying $F \comp H = G\comp H$.
Then, there must exist a unique homomorphism
$M:{\mathcal A}\to {\mathcal E}$
such that $E\comp M = H$.
Let $e_1$ be the element in ${\mathcal E}$ such that 
$M(a_1) = \langle e_1\rangle$ and $E(e_1) = \langle a_1\rangle$.
Since $E$ and $M$ loosely preserve defined functions,
\begin{displaymath}
E(f_2^{\mathcal E}) \subseteq f_2^{\mathcal A} = \langle \bot\rangle
\end{displaymath}
and hence $e_1\notin f_2^{\mathcal E}$, and
$\langle e_1\rangle = M(f_1^{\mathcal A}) \subseteq f_1^{\mathcal E}$.
Therefore, since ${\mathcal E}\in \Mod_\mathrm{CRWL}(T)$, there 
must exist $e_2\in {\mathcal E}$ such that
$e_1 \sqsubset e_2$: otherwise, ${\mathcal E}$ would satisfy 
$f_1\bowtie f_1$ but, since $e_1\notin f_2^\mathcal{E}$, not $\Gamma$.
Besides, due to the monotonicity of $E$ and the equality
$F\comp E = G\comp E$,
it is $E(e_2) = \langle a_1\rangle$.
But then we have $M_1, M_2: {\mathcal B} \to {\mathcal E}$ given by
\begin{displaymath}
\begin{array}{ccc}
M_1(x) = \left\{ \begin{array}{ll}
          \langle\bot\rangle  &\textrm{if $x=\bot$} \\
          \langle e_1\rangle  &\textrm{if $x=b_1$} 
         \end{array} \right. &
\textrm{ and } &
M_2(x) = \left\{ \begin{array}{ll}
          \langle\bot\rangle  &\textrm{if $x=\bot$} \\
          \langle e_2\rangle  &\textrm{if $x=b_1$,} 
         \end{array} \right.
\end{array}
\end{displaymath}
two different homomorphisms satisfying $E\comp M_1 = E\comp M_2$, a 
contradiction with the universal property of equalizers.

In contrast with what happens in CRWL, the following proposition shows a
construction for equalizers in RL.
\begin{proposition}
For all RL-theories $\mathcal{R}=(\Sigma,E,L,\Gamma)$, the category
$\mathcal{R}$-\textup{\textbf{Sys}} has equalizers.
\end{proposition}
\begin{proof}
Let $\mathcal{S}_1$ and $\mathcal{S}_2$ be two $\mathcal{R}$-systems
and let $F,G:\mathcal{S}_1 \to\mathcal{S}_2$ be two
$\mathcal{R}$-homomorphisms between
them; let us build their equalizer
$E: \mathcal{E}\to \mathcal{S}_1$.

The objects in the category $\mathcal{E}$ are those
$s\in\mathcal{S}_1$ such that $F(s)=G(s)$; the arrows, those
$f:s\to s'$ in $\mathcal{S}_1$ such that $F(f)=G(f)$ (which implies,
in particular,
that $F$ and $G$ also coincide over $s$ and $s'$);
composition is that of $\mathcal{S}_1$.
$\mathcal{E}$ is well-defined because functors preserve
identities and composition.

Next, we assign a $(\Sigma,E)$-algebra structure to
$\mathcal{E}$.
For each $f\in\Sigma$ of arity $n$ we define
$f_\mathcal{E}$ to be $f_{\mathcal{S}_1}|_\mathcal{E}$, the
restriction of $f_{\mathcal{S}_1}$ to $\mathcal{E}$.
Let us check that this is a valid definition.
If $e_1,\dots,e_n\in |\mathcal{E}|$ then
\begin{displaymath}
\begin{array}{rcll}
F(f_{\mathcal{S}_1}(e_1,\dots,e_n)) &=
 &f_{\mathcal{S}_2}(F(e_1),\dots,F(e_n))   
 &\quad\textrm{($F$ is homomorphism)} \\
&= &f_{\mathcal{S}_2}(G(e_1),\dots,G(e_n)) 
 &\quad(e_i\in|\mathcal{E}|) \\
&= &G(f_{\mathcal{S}_1}(e_1,\dots,e_n))    
 &\quad\textrm{($G$ is homomorphism)}
\end{array}
\end{displaymath}
and thus $f_{\mathcal{S}_1}(e_1,\dots,e_n)\in |\mathcal{E}|$.
Analogously for arrows.
With this definition it is easy to prove by structural induction
that $t_\mathcal{E} = t_{\mathcal{S}_1}|_{\mathcal{E}^n}$
for all $t(x_1,\dots,x_n)\in T_\Sigma(\mathcal{X})$.
Therefore, for each $t=t'\in E$ it is $t_\mathcal{E} = t_\mathcal{E}'$.

The only thing missing in the definition of $\mathcal{E}$ are
the natural transformations associated to the rewrite rules.
Let
\begin{displaymath}
r:[t(\overline{x})]\to [t'(\overline{x})] \ifcond
  [a_1(\overline{x})]\to [b_1(\overline{x})]\land\dots\land
  [a_m(\overline{x})]\to [b_m(\overline{x})]
\end{displaymath}
be a rule in $\mathcal{R}$.
We have to define a natural transformation
\begin{displaymath}
r_\mathcal{E}: t_\mathcal{E} \comp J_\mathcal{E} \Rightarrow
               t'_\mathcal{E} \comp J_\mathcal{E},
\end{displaymath}
where
$J_\mathcal{E}:\textit{Subeq}
 ((a_{j\mathcal{E}},b_{j\mathcal{E}})_{1\leq j\leq m})\to \mathcal{E}^n$
is the subequalizer functor.
Using the construction of Section~\ref{sec:TMRL} and the fact
that $a_{j\mathcal{E}} = a_{j\mathcal{S}_1}|_{\mathcal{E}^n}$
and $b_{j\mathcal{E}} = b_{j\mathcal{S}_1}|_{\mathcal{E}^n}$,
$1\leq j\leq m$,
it follows that $\textit{Subeq}
 ((a_{j\mathcal{E}},b_{j\mathcal{E}})_{1\leq j\leq m})$
is a subcategory of $\textit{Subeq}
 ((a_{j\mathcal{S}_1},b_{j\mathcal{S}_1})_{1\leq j\leq m})$
and that $J_\mathcal{E}$ is just the restriction of
the corresponding $J_{\mathcal{S}_1}$.
Then we can define $r_\mathcal{E}$ simply by restricting
$r_{\mathcal{S}_1}$, which
is obviously a natural transformation, and this finishes
our construction of $\mathcal{E}$ as an $\mathcal{R}$-system.

Let us now move to the definition of $E$ and the proof that
it is an $\mathcal{R}$-homomorphism.
$E$ is simply the inclusion functor.
If $f\in\Sigma$ and $e_1,\dots,e_n\in |\mathcal{E}|$,
then
\begin{displaymath}
E(f_\mathcal{E}(e_1,\dots,e_n)) = f_\mathcal{E}(e_1,\dots,e_n) =
f_\mathcal{E}(E(e_1),\dots,E(e_n)),
\end{displaymath}
so $E$ is a $\Sigma$-algebra homomorphism.

For a rewrite rule
$r:[t(\overline{x})]\to [t'(\overline{x})] \ifcond
  [a_1(\overline{x})]\to [b_1(\overline{x})]\land\dots\land
  [a_m(\overline{x})]\to [b_m(\overline{x})]$
in $\Gamma$, we have to show that the natural transformation
$E\comp r_\mathcal{E}$ is equal to $r_{\mathcal{S}_1} \comp E^\bullet$.
Let $(\overline{e}^n,\overline{u}^m)\in \textit{Subeq}
 ((a_{j\mathcal{E}},b_{j\mathcal{E}})_{1\leq j\leq m})$.
Regarding $E^\bullet$, we only need to know that
$E^\bullet(\overline{e}^n,\overline{u}^m) =
 (E^n(\overline{e}^n),E^m(\overline{u}^m))$.
Now,
\begin{displaymath}
\begin{array}{rclclcl}
(E\comp r_\mathcal{E})(\overline{e}^n,\overline{u}^m)
&= &E(r_\mathcal{E}(\overline{e}^n,\overline{u}^m)) 
&= &r_\mathcal{E}(\overline{e}^n,\overline{u}^m) \\
&= &r_{\mathcal{S}_1}(\overline{e}^n,\overline{u}^m) 
&= &r_{\mathcal{S}_1}(E^n(\overline{e}^n),E^m(\overline{u}^m)) \\
&= &r_{\mathcal{S}_1}(E^\bullet(\overline{e}^n,\overline{u}^m)) 
&= &(r_{\mathcal{S}_1} \comp E^\bullet)(\overline{e}^n,\overline{u}^m)\,,
\end{array}
\end{displaymath}
so $E$ is an $\mathcal{R}$-homomorphism.

We already know that $\mathcal{E}$ is an $\mathcal{R}$-system and
that $E$ is an $\mathcal{R}$-homomorphism;
the only missing thing is the equalizer property.
Let then $H:\mathcal{C}\to \mathcal{S}_1$ be an
$\mathcal{R}$-homomorphism such that $F\comp H=G\comp H$;
we have to find a unique $M:\mathcal{C}\to \mathcal{E}$
such that $E\comp M= H$.
As $E$ is the inclusion functor the uniqueness is clear,
because the only possibility for all objects $c$ and arrows $u$
in $\mathcal{C}$ is $M(c)= H(c)$ and $M(u)=H(u)$.
It remains to prove that this is a valid definition.
First, because of the equality $F\comp H=G \comp H$
the image of $H$ is
included in $\mathcal{E}$ and $M$ is well-defined;
as $H$ is a functor, so is $M$.
Given $f\in\Sigma$ and $c_1,\dots,c_n\in |\mathcal{C}|$,
we have
\begin{displaymath}
\begin{array}{rcl}
M(f_\mathcal{C}(c_1,\dots,c_n)) &= &H(f_\mathcal{C}(c_1,\dots,c_n)) \\
 &= &f_{\mathcal{S}_1}(H(c_1),\dots,H(c_n)) \\
 &= &f_\mathcal{E}(M(c_1),\dots,M(c_n))
\end{array}
\end{displaymath}
and $M$ is a $\Sigma$-algebra homomorphism.
Finally, if
$r:[t(\overline{x})]\to [t'(\overline{x})] \ifcond
  [a_1(\overline{x})]\to [b_1(\overline{x})]\land\dots\land
  [a_m(\overline{x})]\to [b_m(\overline{x})]$
is a rewrite rule in $\Gamma$ and
$(\overline{c}^n,\overline{u}^m)$ is an object of
$\textit{Subeq}
 ((a_{j\mathcal{C}},b_{j\mathcal{C}})_{1\leq j\leq m})$, then
\begin{displaymath}
\begin{array}{rclcl}
r_\mathcal{E}(M^\bullet(\overline{c}^n,\overline{u}^m))
&= &r_\mathcal{E}(M^n(\overline{c}^n),M^m(\overline{u}^m)) 
&= &r_{\mathcal{S}_1}(H^n(\overline{c}^n),H^m(\overline{u}^m)) \\
&= &r_{\mathcal{S}_1}(H^\bullet(\overline{c}^n,\overline{u}^m)) 
&= &H(r_\mathcal{C}(\overline{c}^n,\overline{u}^m))  \\
&= &M(r_\mathcal{C}(\overline{c}^n,\overline{u}^m))
\end{array}
\end{displaymath}
and we have $M\comp r_\mathcal{C} = r_\mathcal{E} \comp M^\bullet$.
\end{proof}

Note that, in the above proof, the equalizer $\mathcal{E}$ is a model
of all the rewrite rules that $\mathcal{S}_1$ satisfies.
Therefore the result is still valid when $\mathcal{R}$-\textbf{Sys} is 
replaced by the category $\Mod_\mathrm{RL}(\Gamma)$
for some set $\Gamma$ of rewrite rules in the
institution $\mathcal{I}_\mathrm{RL}$.
Since an equivalence of categories preserves limits, we have:
\begin{proposition}\label{institution1:prop}
$\mathcal{I}_\mathrm{CRWL}$ is not embeddable in
$\mathcal{I}_\mathrm{RL}$.
\end{proposition}
\begin{proof}
Let $T$ be the CRWL-theory defined at the beginning of this section.
It has been shown that $\Mod_\mathrm{CRWL}(T)$ does not have all
equalizers, whereas
$\Mod_\mathrm{RL}(\Phi(T))$ has, regardless of the actual
definition of $\Phi$.
Therefore, there cannot exist an equivalence of categories
$\beta_T:\Mod_\mathrm{RL}(\Phi(T)) \to \Mod_\mathrm{CRWL}(\Phi)$.
\end{proof}

What about the other way around?
Can we embed $\mathcal{I}_\mathrm{RL}$ in $\mathcal{I}_\mathrm{CRWL}$?
When we began preparing this work our intuition was that we would be able
to view CRWL as a ``sublogic'' of RL in the first place, but also that
the converse would not be true.
The previous discussion has shown that our intuition was wrong about the
first point and our goal now is to deal with the second.

In order to prove that RL cannot be embedded in CRWL we have to find an
RL-theory $T$ such that $\Mod_\mathrm{RL}(T)$ has a categorical property
that no category of models in CRWL has.
In order to do that, note that for any CRWL-theory $T$ there exists a
CRWL-algebra $\mathcal{A}\in |\Mod_\mathrm{CRWL}(T)|$ with an
infinite number of automorphisms.
Simply consider $\mathcal{A}$ given by
$D^\mathcal{A} = \{\bot,a,b_1,b_2,\dots\}$ with $\bot \sqsubseteq a$,
$\bot \sqsubseteq b_1\sqsubseteq b_2 \sqsubseteq \dots$, the image of
all functions associated to constructor symbols to be $\langle a\rangle$,
and the corresponding one for all defined function symbols to be
$D^\mathcal{A}$.
This way $\mathcal{A}$ is clearly a CRWL-algebra, satisfies all
conditional rewrite rules, and the set
$\{ F_i : \mathcal{A} \to \mathcal{A}\}_{i\in\nat}$, where
\begin{displaymath}
F_i(x)=\left\{ \begin{array}{ll}
               \langle\bot\rangle  & \textrm{if $x=\bot$} \\
               \langle a \rangle   & \textrm{if $x=a$} \\
               \langle b_i \rangle & \textrm{if $x=b_j\ (j\in\nat)$}
               \end{array}
       \right.
\end{displaymath}
is an infinite family of automorphisms of $\mathcal{A}$.
On the other hand, in RL,
if $\mathcal{R}$ is the RL-theory given by
$(\{c\},\{x=c\},\emptyset,\emptyset)$ then, for all $\mathcal{R}$-systems
$\mathcal{S}$, the equality
$\mathit{id}_\mathcal{S} = c_\mathcal{S}$, where $c_\mathcal{S}$ is
a constant functor, forces $\mathcal{S}$ to be a category
with just one object and one arrow, and no infinite family of
homomorphisms can exist.
Therefore (as an equivalence of categories is full and faithful),
$\Mod_\mathrm{RL}(\mathcal{R})$ is not categorically equivalent to
$\Mod_\mathrm{CRWL}(\Phi(\mathcal{R}))$, whatever $\Phi$ might be,
and we have:
\begin{proposition}\label{institution2:prop}
$\mathcal{I}_\mathrm{RL}$ is not embeddable in $\mathcal{I}_\mathrm{CRWL}$.
\end{proposition}

Let us note that Propositions~\ref{institution1:prop} and 
\ref{institution2:prop} still hold even if the general semantics of RL 
is replaced by the preorder semantics mentioned on 
page~\pageref{preorder:pag}.
On the other hand, maps of institutions could be given for the trivial 
semantics in which either everything or nothing can be proven in both logics.

\section{Conclusions}

The main outcome of the research carried out in this paper has been
the clarification of the relationship  between RL and CRWL.
Both logics have been proved to be expressive enough to simulate
deduction in each other in a simple way, though resorting to
binary predicates.
On the other hand, the results on institutions have shown that
neither can RL be considered as a sublogic
of CRWL, nor can CRWL with respect to RL.

During the preparation of this work we have been forced to take
a close look at the notions of entailment system and institution,
and the difficulties we have found have shown us that intuition
can be misleading in this field.
The conclusion we have reached is that it would be very
convenient to develop some kind of generalization of these concepts.
One reason supporting this claim is the fact that, although it
seems clear that CRWL should fit within the framework of entailment 
systems, the lack of the transitivity property forbids it to be 
considered so. 
In addition, there have been several occasions wherein we have had 
to make a distinction between two types of sentences within the same 
logic.
The most outstanding case was that of labeled and unlabeled
rewrite rules in RL, but we should also emphasize that rules in
CRWL-theories are a restricted class of the more general class of
reduction statements.
What all these examples have in common is that sentences belonging
to a theory are given a different treatment from the rest of sentences
and, with the current definitions of entailment system and institution,
there is no way of taking this distinction into account.

Finally, though not presented in the paper due to lack of space, the
results in Section~\ref{rsl:sec} can be used to show that CRWL is 
\emph{reflective} \cite{Palomino-mthesis}. 
Intuitively, this property means that the logic can reason about
itself and has been fruitfully exploited in RL in the design of
programs; thus, an interesting open line of research consists in
the study of ways by which reflection can be exploited in CRWL.

\section*{Acknowledgments}

The author warmly thanks Narciso Mart{\'\i}-Oliet,
Mario Rodr{\'\i}guez-Artalejo, and Jos\'e Meseguer for their 
help in the preparation of this work.

\appendix

\section{Proofs}

\setcounter{proposition}{0}
\begin{proposition}
$\mathcal{E}_\mathrm{RL} = (\Sign, \Sen, \vdash)$ is
an entailment system.
\end{proposition}
\begin{proof*}
The fact that composition of signature morphisms is associative
(for equational logics in general, and for our unsorted and
unconditional case in particular) is all that is needed to check
that \Sign\ is a category and \Sen\ a functor.
Regarding the properties that $\vdash$ must satisfy:
\begin{enumerate}[4.]
\item reflexivity: By \textbf{Replacement}
 (combined with \textbf{Implication introduction} for conditional rules).
\item monotonicity: Immediate by the definition of the entailment 
 relation.
\item transitivity: Assume $\Gamma\vdash \varphi_i$ for all $i\in I$ 
 and $\Gamma \cup \{\varphi_i \mid i\in I\} \vdash \psi$.
 The easiest way to prove $\Gamma \vdash \psi$ is by resorting
 to the soundness and completeness of the RL-calculus.
 Let $\mathcal{S}$ be a $\Gamma$-system, so 
 $\mathcal{S}\models \varphi_i$ for all $i\in I$.
 Therefore $\Gamma$ can also be considered a
 $\Gamma\cup \{\varphi_i \mid i\in I\}$-system and then
 $\mathcal{S} \models \psi$.
\item $\vdash$-translation: Suppose $\Gamma \vdash \varphi$.
 Given a theory morphism $H$, it can be proved by induction on the
 derivation that $\Sen(H)(\Gamma) \vdash \Sen(H)(\varphi)$.
 The only non-trivial case is the one corresponding to 
 \textbf{Replacement} and we illustrate it with an unconditional 
 rule.
 If, for some $[t(\overline{x})] \to [t'(\overline{x})] \in \Gamma$,
 the last step in the derivation of $\Gamma\vdash \varphi$ is 
 \begin{displaymath}
 \bigfrac{[w_1]\to [w'_1] \;\;\; \dots \;\;\;
          [w_n] \to [w'_n]}
         {[t(\overline{w}/\overline{x})] \to
          [t'(\overline{w'}/\overline{x})]}
 \end{displaymath}
 then, by the induction hypothesis,
 $\Sen(H)(\Gamma) \vdash \Sen(H)([w_i] \to [w'_i])$
 for $i=1,\dots,n$, and, since 
 $\Sen(H)([t]\to [t']) = [H(t)] \to [H(t')]$
 belongs to $\Sen(H)(\Gamma)$, we can build a derivation for
 $\Sen(H)([t(\overline{w}/\overline{x})] \to
          [t'(\overline{w'}/\overline{x})])$
 from $\Sen(H)(\Gamma)$
 by applying \textbf{Replacement}.$\mathproofbox$
\end{enumerate}
\end{proof*}

\setcounter{proposition}{5}
\begin{proposition}
Given any RL-theory $T=(\Sigma,E,\Gamma)$, and
$l,r \in T_\Sigma(\mathcal{X})$:
\begin{displaymath}
\begin{array}{rcl}
T\vdash_\mathrm{RL} [l]\to[r] &\iff
 &(\exists l'\in[l],\exists r'\in [r]) \;\;
 \beta(T)\vdash_\mathrm{CRWL} R(l',r')\to\True  \\
&\iff &(\forall l'\in[l],\forall r'\in [r]) \;\;
 \beta(T)\vdash_\mathrm{CRWL} R(l',r')\to\True
\end{array}
\end{displaymath}
\end{proposition}
\begin{proof}
By Propositions~\ref{prop:RLinCRWLhalf} and \ref{prop:RLinCRWLmainhalf},
it is enough to see that if
$(\Sigma\cup\{\True\},E,\Gamma)\vdash_\mathrm{RL} [l]\to [r]$
then
$(\Sigma,E,\Gamma)\vdash_\mathrm{RL} [l]\to [r]$.
The easiest way of proving this implication is by using the
completeness of the RL-calculus.

Note that, since $\True$ does not belong to $\Sigma$ (and hence
it appears neither in $E$ nor in $\Gamma$), a model of 
$(\Sigma\cup\{\True\},E,\Gamma)$ is just a model of
$(\Sigma,E,\Gamma)$ together with an interpretation for the
constant $\True$, and therefore either both satisfy
$[l]\to [r]$ or none does.
But then
\begin{displaymath}
\begin{array}{rcl}
(\Sigma,E,\Gamma) \vdash [l] \to [r] &\iff
&(\Sigma,E,\Gamma) \models [l] \to [r] \\
&\iff &(\Sigma\cup\{\True\},E,\Gamma) \models [l] \to [r] \\
&\iff &(\Sigma\cup\{\True\},E,\Gamma) \vdash [l] \to [r] 
\end{array}
\end{displaymath}
whence $(\Sigma,E,\Gamma) \vdash [l] \to [r]$ follows.
\end{proof}

\end{document}